\documentclass[longauth]{aa} 

\usepackage{graphicx}
\usepackage{txfonts}
\usepackage{xcolor} 
\usepackage{hyperref}
\hypersetup{
    colorlinks = True,
    citecolor=blue,
    linkcolor = cyan,
}

\begin{document} 

\newcommand{\ngc}{NGC\,4388}

\newcommand{\red}[1]{\textcolor{red}{#1}}
\newcommand{\green}[1]{\textcolor{olive}{#1}}

\newcommand{\oiii}{[\ion{O}{III}]$\lambda5007\AA$\xspace}
\newcommand{\halpha}{H$\alpha$\xspace}
\newcommand{\hbeta}{H$\beta$\xspace}
\newcommand{\nii}{[\ion{N}{II}]$\lambda6583\AA$\xspace}
\newcommand{\siiALL}{[\ion{S}{II}]$\lambda\lambda6716,6731$\xspace}
\newcommand{\siia}{[\ion{S}{II}]$\lambda6716$\xspace}
\newcommand{\siib}{[\ion{S}{II}]$\lambda6731$\xspace}

\newcommand{\moka}{\ensuremath{\mathrm{MOKA^{3D}}}\xspace}

\newcommand{\kms}{\ensuremath{\mathrm{km\,s^{-1}}}\xspace}
\newcommand{\msun}{\ensuremath{{M_\odot}}\xspace}

\newcommand{\MVZ}[1]{\textcolor{magenta}{#1}}

\title{Low-power jet-ISM coupling and disk-plane perturbations in the edge-on Seyfert galaxy NGC 4388}

\author{M.~Ceci \inst{\ref{UNIFI},\ref{arcetri}}
\and G.~Venturi \inst{\ref{arcetri},\ref{SNS}}
\and C.~Marconcini \inst{\ref{arcetri}}
\and M.~V.~Zanchettin\inst{\ref{arcetri}}
\and G.~Cresci \inst{\ref{arcetri}}
\and M.~Tart\.{e}nas \inst{\ref{FTMC}}
\and K.~Zubovas \inst{\ref{FTMC}}
\and E.~Treister \inst{\ref{UTA}}
\and E.~Amenta \inst{\ref{UNIBO},\ref{IRA}}
\and E.~Bertola \inst{\ref{arcetri}}
\and F.~Belfiore \inst{\ref{ESO},\ref{arcetri}}
\and C.~Bracci \inst{\ref{UNIFI},\ref{arcetri}}
\and E.~Cataldi \inst{\ref{UNIFI},\ref{arcetri}}
\and Q.~D'Amato \inst{\ref{IAPS}}
\and A.~Feltre \inst{\ref{arcetri}}
\and M.~Ginolfi \inst{\ref{UNIFI},\ref{arcetri}}
\and I.~Lamperti \inst{\ref{CAB}}
\and F.~Mannucci \inst{\ref{arcetri}}
\and A.~Marconi \inst{\ref{UNIFI},\ref{arcetri}}
\and A.~Marasco \inst{\ref{OAPd}}
\and M.~Meenakshi \inst{\ref{Leibniz}}
\and M.~Mingozzi \inst{\ref{STScI}}
\and B.~Moreschini \inst{\ref{UNIFI},\ref{arcetri}}
\and D.~Mukherjee \inst{\ref{IUCAA}}
\and E.~Nardini \inst{\ref{arcetri}}
\and M.~Perna \inst{\ref{CAB}}
\and F.~Ricci \inst{\ref{RomaTre}}
\and M.~Scialpi \inst{\ref{unitn},\ref{UNIFI},\ref{arcetri}}
\and L.~Ulivi \inst{\ref{CAB}}
    }

\institute{
\label{UNIFI} Università di Firenze, Dipartimento di Fisica e Astronomia, via G. Sansone 1, 50019 Sesto Fiorentino, Florence, Italy 
\and
\label{arcetri} INAF - Arcetri Astrophysical Observatory, Largo E. Fermi 5, I-50125, Florence, Italy 
\and
\label{SNS} Scuola Normale Superiore, Piazza dei Cavalieri 7, 56126 Pisa, Italy
\and
\label{FTMC} Center for Physical Sciences and Technology, Saulėtekio al. 3, Vilnius LT-10257, Lithuania
\and
\label{UTA} Instituto de Alta Investigaci\'on, Universidad de Tarapac\'a, Casilla 7D, Arica, Chile
\and
\label{UNIBO} Dipartimento di Fisica e Astronomia, Università di Bologna, Via P. Gobetti 93/2, 40129 Bologna, Italy
\and
\label{IRA} INAF – Istituto di Radioastronomia, Via P. Gobetti 101, 40129 Bologna, Italy
\and
\label{ESO} European Southern Observatory, Karl-Schwarzschild Stra$\ss$e 2, D-85748 Garching bei M\"unchen, Germany
\and
\label{IAPS} INAF - Istituto di Astrofisica e Planetologia Spaziali, Via Fosso del Cavaliere 100, 00133 Rome, Italy
\and
\label{OAPd} INAF - Osservatorio Astronomico di Padova, Vicolo dell’Osservatorio 5, I-35122 Padova, Italy
\and
\label{Leibniz} Leibniz Institute for Astrophysics, An der Sternwarte 16, D-14482 Potsdam, Germany
\and
\label{STScI} AURA for ESA, Space Telescope Science Institute, 3700 San Martin Drive, Baltimore, MD 21218, USA
\and
\label{IUCAA} IUCAA, Post Bag-4, Pune University, Ganeshkhind, Pune 411007, India
\and
\label{CAB} Centro de Astrobiología (CAB), CSIC–INTA, Cra. de Ajalvir Km. 4, 28850 – Torrejón de Ardoz, Madrid, Spain
\and
\label{RomaTre} Dipartimento di Matematica e Fisica, Università degli Studi di Roma Tre, via della Vasca Navale, 84, 00146, Roma, Italy
\and
\label{unitn} University of Trento, Via Sommarive 14, I-38123 Trento, Italy
}
    
\date{Received MM dd, year; accepted MM dd, year}
\abstract
{Active galactic nuclei (AGN) are thought to regulate galaxy evolution through feedback processes, including radiatively-driven winds and collimated jets. However, their relative contribution and their efficiency in coupling with the interstellar medium (ISM) remain poorly constrained, even in nearby systems.}
{We investigate the ionized gas in the nearby ($D$~$\sim$~17~Mpc) Seyfert~2 galaxy \ngc, which hosts a galaxy-scale ($\sim$~1.6~kpc), low-kinetic-power ($\sim$~1.4$~\times~10^{42}$~erg~s$^{-1}$) bipolar radio jet that appears, in projection, roughly perpendicular to the edge-on galaxy disk plane.
In particular, we assess how AGN-driven outflows and radio activity affect the host-galaxy ionized ISM.}
{We analyze optical integral--field spectroscopic observations obtained with the Wide Field Mode with adaptive optics of VLT/MUSE, complemented by VLA radio data tracing the jet. We produce spatially resolved maps of ionized gas emission, kinematics, excitation, and dust attenuation. We reconstruct the de-projected 3D outflow geometry and velocity field using the \moka\ kinematic model, and estimate its mass and energetics.}
{The ionized gas shows a rotating disk coexisting with a kpc-scale biconical outflow. We measure a total ionized outflow mass of $\sim10^6$~M$_\odot$, a mass outflow rate of $\sim0.2$~M$_\odot$~yr$^{-1}$, and a kinetic power of $\sim3\times10^{40}$~erg~s$^{-1}$, which gives a coupling efficiency of $\sim$1$\%$ relative to both the jet kinetic and the radiation-driven nuclear wind power, indicating that either mechanism could drive the outflow. In addition to the outflow cones oriented perpendicular to the galaxy disk, we detect an extended ($\sim2$~kpc) enhancement of the gas velocity dispersion ($\sigma$) along the galaxy disk plane, oriented perpendicular to the outflow and jet axes. These regions show optical line ratios consistent with shock excitation. ALMA CO(2--1) data show only a marginal spatial correspondence with this $\sigma$-enhancement and reveal not clear evidences of perturbed high velocity gas. Overall, these features can be interpreted as perturbations induced by the interaction between the radio jet and the host galaxy disk ISM, although the co-spatial enhancement in stellar velocity dispersion precludes attributing them exclusively to this mechanism.}
{
Our results support a scenario in which relatively low-power radio structures, even oriented nearly perpendicular to the disk, could efficiently couple with the ISM and inject significant energy. This is the first observational case of such coupling in this orientation regime and suggests that jet-induced perturbations represent an important and possibly ubiquitous channel of AGN feedback that should be considered independently of jet orientation.}

\maketitle
%

\section{Introduction}
Feedback from active galactic nuclei (AGN) has become a key ingredient in models of galaxy evolution, as it provides a natural explanation for the relations observable between the host galaxy and the supermassive black hole (SMBH) properties \citep[e.g.,][]{Kormendy+1995,Haring2004,Marconi2004,Ferrarese2000,Gebhardt2000,Fabian2012}.
AGN feedback is traditionally believed to operate in two main modes.
In the radiative mode (also quasar/wind mode) radiation pressure drives powerful outflows that can suppress star formation (SF) by removing, displacing, or heating the host galaxy interstellar medium \citep[ISM; e.g.,][]{Cicone2014,Harrison14,Cresci2015,Carniani2015,DallAgnoldeOliveira2021}.
In the kinetic mode (also radio/jet mode) powerful ($\ge10^{45}$ erg s$^{-1}$), large-scale ($\sim$100 kpc) collimated jets launched from the AGN heat the circumgalactic medium \citep{Fabian2012}, thereby quenching SF in massive galaxies at the center of cool-core clusters \citep[e.g.,][]{McNamara2012}. 
Recently, both observations \citep{Cresci2015,Molyneux2019,Venturi2021_turmoil,Venturi2023,Girdhar2022,PeraltadeArriba2023,Ulivi2024} and theoretical studies \citep{Mukherjee2018a,Mukherjee2016,Mukherjee2018b,Mukherjee2025,Mandal2021,Meenakshi2022} have shown that even low-power ($<10^{44}$ erg s$^{-1}$), compact ($\leq1$ kpc) jets, now resolvable on sub--kpc scales, can strongly affect the ISM, injecting perturbations, accelerating outflows, and heating the gas reservoir for SF \citep[e.g.,][]{Oosterloo2017,Dasyra2016,Dasyra2022,Audibert2023,Dasyra2024}, suggesting that the classical radiative-versus-kinetic dichotomy may not capture the full diversity of feedback phenomena \citep[e.g.,][]{Heckman2014,HarrisonRamosAlmeida2024}. A compelling example is \citet{Venturi2021_turmoil}, who found that in local Seyferts, low-power compact radio jets, with low inclination angle ($\lesssim45^{\circ}$) to the disk plane, produce a systematic broadening of ionized-gas lines on kpc scales. Such effect is observed perpendicularly to the jet axis and without a coherent velocity shift, consistent with jet-driven turbulence.

This picture is supported by statistical studies on large AGN samples \citep{Mullaney2013, Molyneux2019}, which show that moderate radio luminosities (L$_{1.4\,\text{GHz}} \approx 10^{23}-10^{25}$ W Hz$^{-1}$, corresponding to $\nu L_{
\nu} \approx 10^{39}-10^{41}$ erg s$^{-1}$) correlate with significant broadening of the [\ion{O}{III}] emission-line profile, consistent with the effects of low-power jet activity (see also \citealt{Wen2026}).
Recently, an enhancement in gas velocity dispersion ($\sigma$) relative to rotational motion has been reported perpendicular to the radio jet axis has been reported in several AGN systems, spanning a wide range of bolometric luminosities, both in ionized gas \citep[e.g.,][]{Venturi2023,Ulivi2024,Speranza2024} and in the cold molecular phase \citep[e.g.,][]{Girdhar2022,Audibert2023} of radiatively powerful AGN (L$_\text{bol}~\geq~10^{45}$~erg~s$^{-1}$), indicating that even low-power jets can strongly perturb the inner ISM \citep[see also][]{RamosAlmeida2022,Audibert2025}. Notably, \citet{Ulivi2024} found that in two of four type-2 AGN with moderately powerful radio emission ($\sim10^{44}$ erg s$^{-1}$) and a clearly detected jet, the velocity dispersion increases up to 1000~\kms\ perpendicular to the jet.
Similar signatures have been reported in nearby Seyferts (see \citealt{Mukherjee2025} for a review), both ionized \citep[e.g.,][]{Couto2017, Finlez2018, RuschelDutra2021} and molecular \citep[e.g.,][]{Riffel2015, Shimizu2019, Bianchin2021}, though their physical link to the jet remains debated, with alternatives invoking beam smearing, equatorial disk winds, or multi-directional outflows (see \citealt{Couto2023}).
Support for a jet-driven origin has nonetheless grown through numerical simulations showing that jet-ISM coupling is regulated by jet power and inclination, with low-inclination, low-power jets driving the strongest disk-ISM perturbations \citep{Mukherjee2018a}.

However, the physical origin of the observed $\sigma$-enhancement perpendicular to the jet axis and the jet-ISM coupling remains debated. Nearby Seyfert galaxies provide a unique opportunity to probe these processes on a spatially resolved basis down to scales of hundreds or even tens of pc. \ngc\ represents an ideal laboratory for this purpose.
The target is a local radio-quiet Seyfert 2 galaxy in the Virgo cluster (with a redshift of $z$~=~0.008419\footnote{From the NASA Extragalactic Database (\hyperlink{https://ned.ipac.caltech.edu/}{NED}).}) with a well-known ionized outflow and clear signatures of radio emission on sub-kiloparsec scales \citep{Helou1981}.
The disk is strongly inclined ($\sim$78$^{\circ}$)\footnote{Throughout this paper, we adopt the convention where an inclination of $0^{\circ}$ corresponds to a face-on disk and $90^{\circ}$ to an edge-on disk.} along the east-west (E-W) direction on the sky, with its northern edge closest to the observer \citep{Pogge1988, Cayatte1990, Veilleux1999a}. Evidence for bar streaming in the disk has also been reported, with a projected position angle (PA) of $\sim100^{\circ}$ on the sky \citep{Veilleux1999a}.
\ngc\ hosts a prominent biconical Narrow Line Region (NLR) extending several kiloparsecs above and below the disk along the northeast-southwest (NE-SW) direction, first identified by \citet{Pogge1988,Yoshida2002} with an opening angle of $\sim90^{\circ}$ \citep{Schmitt2003a,Schmitt2003b} The [O~\textsc{iii}] emission is dominated by the southern cone, while the northern counter-cone is partially obscured by the galaxy disk. 
Ionized gas is spatially coincident with soft X-ray emission, indicating a common origin \citep{Matt1994, Veilleux1999a, Iwasawa2003, Bianchi2006}.  A large-scale H\,{\sc i} tail extends up to $\sim110$~kpc towards NE, likely associated with an ionized gas plume, has also been detected \citep{Pogge1988, Corbin1988, Colina1992, Falcke1998, Veilleux1999b,Yoshida2004,Oosterloo2005}.
\ngc\ shows one of the strongest nuclear molecular gas deficits in The Galaxy Activity, Torus, and Outflow Survey (GATOS) sample, consistent with efficient nuclear AGN feedback. The molecular gas rotates in a highly inclined ($\sim79^\circ$) ring with a deprojected radius of $r \sim 100$, co-spatial with the stellar disk \citep{DominguezFernandez2020, GarciaBurillo2021}. The ring has a clumpy structure, with pronounced minima in molecular-gas emission aligned with the jet axis (PA $\sim15^\circ$--$20^\circ$), suggesting preferential gas depletion along the jet direction.

Near-IR IFU data revealed a nuclear disk and ionized gas aligned with the NLR and radio axis on scales of $\sim$100--200~pc, suggesting a close link between nuclear structures and outflow activity \citep{Greene2014, RodriguezArdila2017}. 
Recent JWST/MIRI mid-IR imaging \citep{Haidar2026, Rosario2026} revealed extended hot dust emission reaching up to $\sim$150 pc from the nucleus, closely tracing both the NLR and radio jet morphology. AGN photoionization alone is insufficient to explain the observed excess MIR emission, which instead requires an additional contribution from shock heating driven by jet-ISM interaction.

\begin{figure*}[h!]
    \centering
    \includegraphics[width=0.99\linewidth, trim=13 10 6 10, clip]{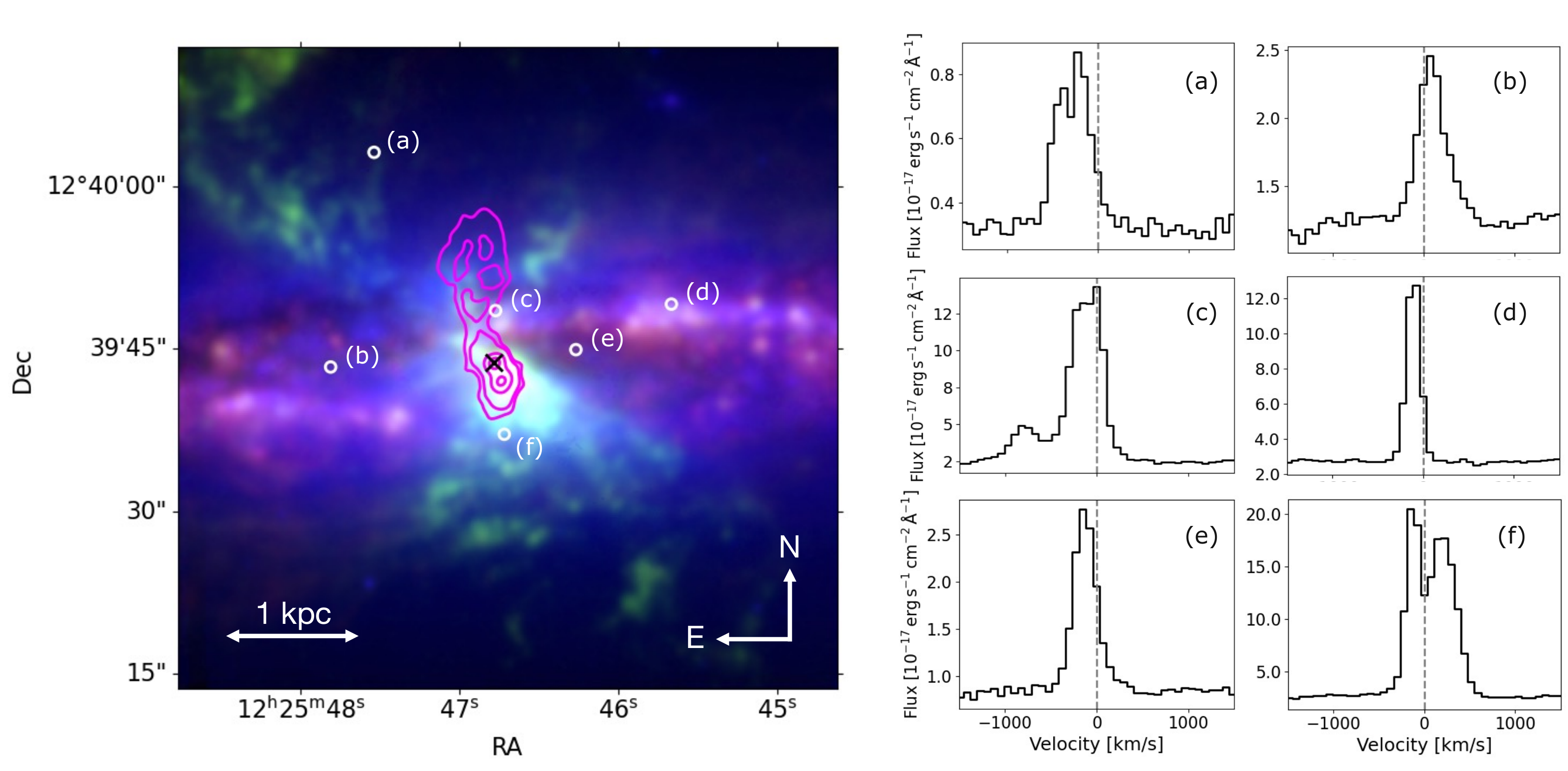}
    \caption{Three-color composite image of \ngc\ and [O~\textsc{iii}] line profiles. The left panel shows the MUSE RGB image, where green, red, and blue correspond to \oiii, \halpha, and the stellar continuum obtained by collapsing the MUSE data cube over the wavelength range $\sim$4780--6570\,\AA, respectively. Contours show the $(5,\,10,\,50,\,120)~\times~$rms levels of the VLA 6\,GHz image. The cross marks the position of the nucleus from the VLA 6 GHz image. White circles labelled (a)-(f) indicate the 0.4\arcsec-radius apertures used to extract the spectra shown in the right panels. The right panels display the \oiii line profiles at the corresponding locations, plotted as a function of velocity relative to the systemic redshift (vertical dashed line).}
    \label{fig: rgb image}    
\end{figure*}

In hard X-rays, \ngc\ is one of the brightest local Seyferts. 
Combining the BAT AGN Spectroscopic Survey (BASS; \citealp{Koss17,Koss22,Ricci17}) bolometric luminosity (L$_\text{bol}~\simeq~$9$\times$10$^{43}$~erg~s$^{-1}$; \citealt{Koss22}) with the maser-based black hole mass (M$_\text{BH}$~=~8.5$\times$10$^{6}$~M$_\odot$; \citealt{Kuo11}) gives an Eddington ratio $\lambda_\text{Edd}~\simeq~$0.08.

In the radio band, studies by \cite{Stone1988} and \cite{Hummel1991} identified two spatial peaks, a primary one at the nucleus and a secondary one located 230 pc to the southwest. Additionally, a plume of radio plasma extends northward from the optical nucleus. The strong correlation between optical line features and radio morphology \citep{Falcke1998} made \ngc\ a textbook case of jet-ISM interaction.
However, \citet{Sargent2024} recently suggested that this morphology may instead reflects wind-driven shocks.  
Indeed, the alignment of the southwest component with soft X-ray ($\sim$0.5--2~keV) and [\ion{O}{III}] emission along the ionization cone, combined with its steep spectrum, and strong [\ion{Fe}{II}] emission, recreate a scenario more consistent with wind-driven shocks than a collimated relativistic jet. Regardless of the driving mechanism, the co-spatial alignment of radio, X-ray, and ionized-gas emission indicates strong coupling between the radio plasma and the ISM. A similar configuration, interpreted in the context of jet-ISM interaction, has recently been reported in NGC~4438 by \cite{PuigSubira2026}. Nonetheless, for the purposes of the following analysis, we adopt the working assumption that the radio structure traces a relativistic jet.

Despite this extensive characterization of \ngc, it remains untested whether its low-power radio jet, oriented nearly perpendicular to the disk, can still drive the disk-plane $\sigma$-enhancement found in other Seyferts with low-inclination jets. Here we combine new VLT/MUSE optical IFU data with VLA radio and ALMA CO(2--1) observations to investigate this question.
This paper is organized as follows. In Sect. \ref{sec:obs, data redutction, data analysis}, we introduce the optical and radio observations and we summarize the MUSE data analysis. In Sect. \ref{sec: results} we discuss our results through the characterization of the ISM properties and the spatially resolved kinematics of the ionized gas in the central region of \ngc. Finally, in Sect. \ref{sec: summary}, we summarize our results. For all the maps shown in this work, north is up and east is to the left.

\section{Overview of observations and data analysis}\label{sec:obs, data redutction, data analysis}
In this work we combine optical integral-field spectroscopy of \ngc\ obtained with  MUSE \citep{Bacon2010} at the Very Large Telescope (VLT) in Wide Field Mode (WFM) with adaptive optics (AO), covering the central $\sim5\times5$ kpc$^2$ at an effective spatial resolution of $\sim0.8\arcsec$ (corresponding to $\sim~65$~pc at the distance of 16.8 Mpc; \citealp{Merloni2003}). We complement these data with archival Karl G. Jansky Very Large Array (VLA) C- and K-band (6 and 20 GHz) continuum maps, with synthesized beams of $\sim1.3\times0.9$ and $\sim1.6\times1.2$ arcsec$^2$, and with ALMA CO observations tracing the diffuse cold molecular gas emission.
A detailed description of the observations and data reduction is provided in Appendix~\ref{app: obs, data reduction}.

In this paper, we adopt as the nucleus position the northern peak in the VLA 6 GHz image (see Fig.~\ref{fig: radio images}), that is RA~=~$12^{\rm h}25^{\rm m}46.786^{\rm s}$,  Dec~=~$+12^\circ39'43.711''$.

The left panel of Fig.~\ref{fig: rgb image} shows a three-color composite image of \ngc, including \oiii, \halpha, and the stellar continuum derived from the MUSE cube collapsed over $\sim$4780--6570 \AA, after masking emission lines and sky residuals. The \oiii\ emission traces the biconical NLR outflow in the N-S direction, extending toward the NE ionized-gas plume observed by \citet{Yoshida2002}, while \halpha\ traces SF in the disk, revealing several star-forming clumps. The blue continuum emission highlights the SF regions likely tracing spiral arms in upper-right and lower-left sides of the disk, with darker regions marking dust lanes.
Additionally, the VLA 6 GHz contours tracing galaxy-scale ($\sim1.6$~kpc) bipolar radio jet further confirm the correlation between outflow and radio morphology noted by \citet{Falcke1998}.
The right panels of Fig.~\ref{fig: rgb image} show the [O~\textsc{iii}] line profiles extracted at different locations across the FoV: multiple spectral peaks appear across the outflow and plume (apertures a, c, f), while disk spectra (apertures b, d, e) show a range of line widths (see Sect.~\ref{sec: moment maps}).

For the analysis of MUSE data, we followed the methodology of \cite{Marasco2020} and \cite{Tozzi2021}, applying a two-step procedure to the MUSE cube. First, we modeled and subtracted the stellar continuum using Voronoi-binned spectra and \textsc{pPXF} \citep{cappellari_adaptive_2003, cappellari_parametric_2004}, fitting the stellar kinematics with E-MILES templates \citep{Vazdekis2016} and masking wavelength ranges affected by instrumental artifacts (see Appendix~\ref{app: stellar fit}). Second, on the continuum-subtracted cube, we performed a spaxel-by-spaxel multi-Gaussian fit (up to three components) of the main emission lines, using the stellar kinematics as a prior for the disk component where appropriate and selecting the optimal number of components via a Kolmogorov–Smirnov test on the residuals (full details in Appendix~\ref{app: data analysis}).

\section{Results} \label{sec: results}
In this section, we present the main results of our analysis of the ionized gas in \ngc. We first discuss the stellar kinematics in Sect.~\ref{sec: stellar kinematics} and examine the morphology and kinematics of the ionized gas through the moment maps in Sect.~\ref{sec: moment maps}. Then, in Sect.~\ref{sec: attenuation and BPT} we discuss the dust attenuation and ionization properties. Finally, we model the outflow kinematics and use the resulting constraints to estimate the outflow energetics and evaluate the impact of the radio jet on the host galaxy ISM (see Sect.~\ref{sec: energetics}). 

\subsection{Stellar kinematics} \label{sec: stellar kinematics}

The last column of Fig.~\ref{fig: moment maps} shows the velocity and velocity dispersion maps of the stellar component from the pPXF fitting.
Voronoi bins with more than 2000 spaxels were masked, as they correspond to outer regions where the stellar disk is undetected (see continuum map in Fig.~\ref{fig: rgb image}).
The stellar velocity map (center-right panel of Fig. \ref{fig: moment maps}) reveals disk rotation, with receding velocities of $\sim$120 \kms toward the east and approaching velocities of $\sim$~--130 \kms toward the west. The zero-velocity curve exhibits a twisted morphology, typically indicating a stellar bar. This is consistent with \citet{Veilleux1999a}, who inferred a large-scale bar (deprojected radius $\sim$1.5~kpc, PA~$\sim100^\circ$) from the H$\alpha$ velocity field (see also \citealt{DominguezFernandez2020}).
An enhancement in velocity amplitude is found $\sim$2\arcsec\ from the nucleus on both sides of the disk. The elongated NE-SW structure connecting these enhancements is spatially coincident with the nuclear disk found by \citet{Greene2014} (PA~$\sim75^\circ$), also visible in the continuum color map (inset in Fig.~\ref{fig: attenuation map}, lower panel).
The bottom-right panel of Fig. \ref{fig: moment maps} shows a velocity dispersion pattern broadly consistent with a rotating disk, with lower $\sigma$ ($\sim$40 \kms) in the outer regions and a central enhancement reaching $\sim$110 \kms. Using near-IR IFU spectroscopy, \citet{Greene2014} and \citet{RodriguezArdila2017} instead found a $\sigma$-drop of $\sim$50 \kms\ at 150~pc NE of the center, attributed to a disk within the inner 180~pc embedded in the larger-scale bulge/bar. From our data, we do not detect such variation, likely due to differences in the wavelength ranges probed and dust attenuation.

\subsection{Moment maps} \label{sec: moment maps}

\begin{figure*}[t]
    \centering
    \includegraphics[width=1\linewidth, trim=7 13 12 10, clip]{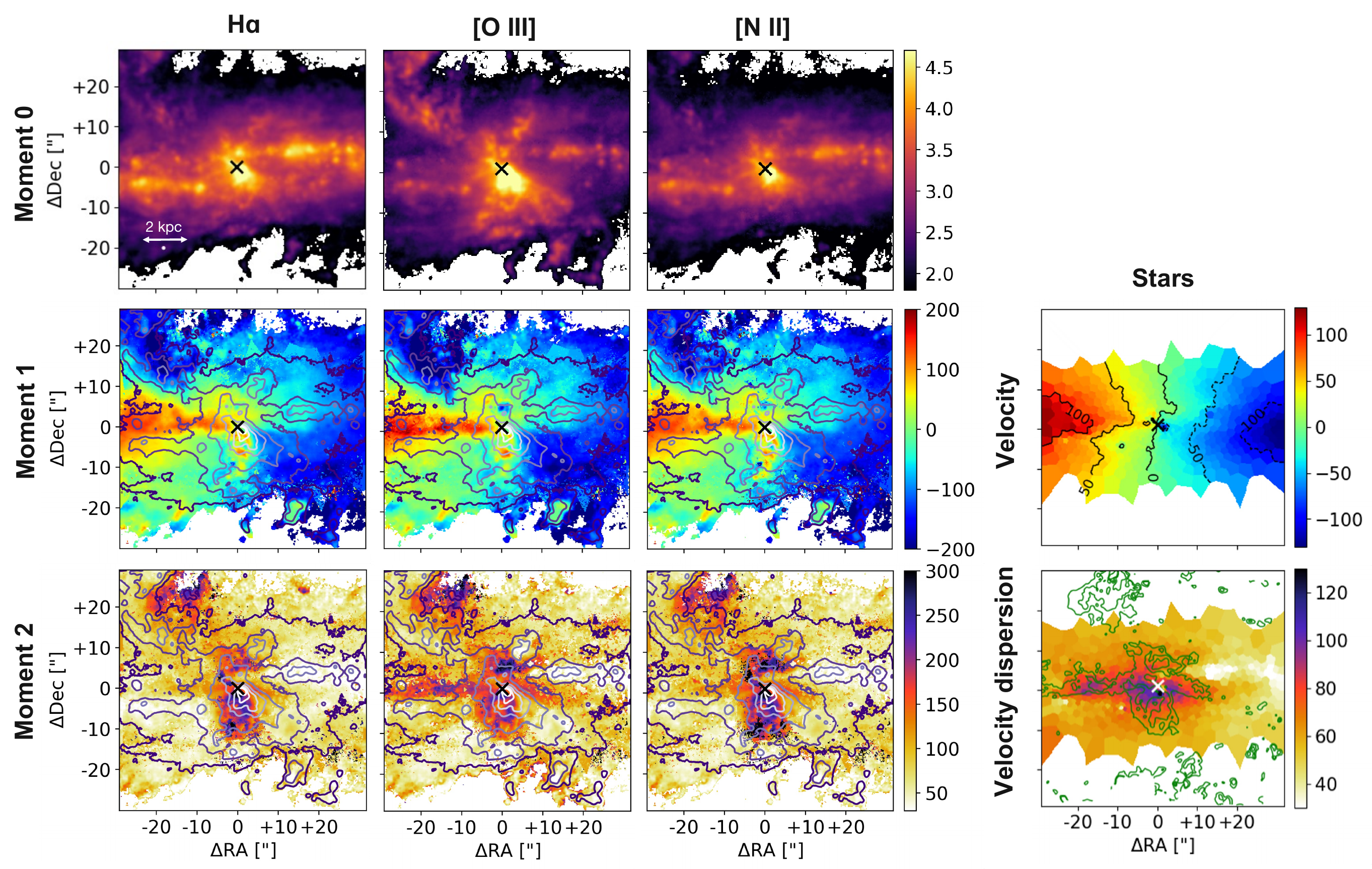}
    \caption{Stellar kinematics and moment maps of the ionized gas in \ngc. 
    The first three columns show the \halpha, \oiii, and \nii\ emission-line moment maps (see Sect.~\ref{app: lines fitting} for details). From top to bottom, we show the flux (in logarithm scale, in units of 10$^{-17}$ erg s$^{-1}$ cm$^{-2}$), the LOS velocity (in \kms), and the velocity dispersion map (in \kms) of the total modeled line profile. The contours overlaid on the moment 1 and moment 2 maps correspond to (40\%, 50\%, 60\%, 70\%, 80\%, 90\%, and 95\%) of the peak \oiii\ flux.
    Spaxels with S/N $\leq$ 3 are masked.       
    The last column displays the stellar velocity and velocity dispersion derived with the pPXF fitting (both in \kms; see Sect.~\ref{app: stellar fit} for details).
    Black contours represent the stellar velocity values as indicated. White contours represent the (120, 150, 200, 250)~\kms levels of the \oiii second-moment map. Voronoi bins containing more than 2000 spaxels are masked out.
    In the \halpha flux map, the white circle represent the MUSE PSF ($\sim0.9''$). 
    The cross marks the position of the nucleus from the VLA 6~GHz image. 
    }
    \label{fig: moment maps}
\end{figure*}

In Fig.~\ref{fig: moment maps} we present the moment maps of \oiii, \halpha, and \nii, obtained from our fitting procedure. We show the observed flux maps, without correcting them for dust attenuation, since the attenuation correction is derived from the Balmer decrement (\halpha/\hbeta) and the low S/N of \hbeta limits the usable FoV (see Sect. \ref{sec: attenuation and BPT} and Fig. \ref{fig: attenuation map}).
For all species, the zeroth-moment maps trace gas emission from both the galaxy disk and the ionization cones. 
While \halpha primarily trace the disk, the \oiii\ and \nii\ emission lines highlight the NLR shape. 
In particular, the \oiii\ map clearly outlines the edge of the two cones. 
For all the fitted species, the southern cone is located southwest of the nucleus. When accounting for attenuation (see Sect.~\ref{sec: attenuation and BPT}), we recover the same spatial distribution reported by \cite{Greene2014} for the Br$\gamma$ and [\ion{Si}{VI}] emission lines. This implies that, for the first time, we are able to observe also the northern ionization cone in the optical at these nuclear scales ($\sim$100--200~pc).

From the first-moment maps, we identify different kinematic features. In the E-W direction, the same rotating pattern observed for the stars in Fig. \ref{fig: moment maps} is present, with absolute velocities higher than $\sim$50 \kms, underlying the slightly different rotating velocities between the gas and the stars. 
The ionized gas plume in the NE of the FoV is characterized by approaching velocities ($<$ --300 \kms). These values are more negative than the ones obtained by \cite{Veilleux1999b}, probably due to the higher sensitivity of MUSE, which allows us to detect fainter emission.
We observe complex velocity structures in the inner regions of the biconical outflow, particularly in the \oiii\ map, which provides the clearest tracer of the outflow. As an example, at 5.5\arcsec\ to the north (south) of the nucleus, a clumpy region with a velocity of --170 \kms (+150 \kms) is present. These and other features are tracing different kinematical components of the biconical outflow (see Sect. \ref{sec: moka fit results}). Unlike the north cone, on large scales the south cone has a velocity gradient in the direction perpendicular to the outflow axis. In fact, a clear velocity gradient is observed across the cone along the east-west direction, from $\sim+150$~\kms\ on the eastern side to $\sim-200$~\kms\ on the western side, passing through $\sim0$~\kms\ near the cone axis. This gradient, perpendicular to the outflow axis, is suggestive of rotation inherited from the disk/circumnuclear gas from which the outflowing material originates, possibly combined with drag from the surrounding ISM.
We find a more complex structure in the nuclear region compared to the velocity maps of Br$\gamma$ and [\ion{Si}{VI}] presented by \cite{Greene2014}. This is likely due to the higher sensitivity of MUSE to diffuse gas compared to SINFONI, as well as the stronger dependence on dust attenuation in the optical regime, which is particularly high in the nuclear region (see Fig.~\ref{fig: attenuation map}).

From the velocity dispersion maps in Fig.~\ref{fig: moment maps}, we identify complex structures in both the ionization cone and the galaxy disk. In the former, a $\sigma$-enhancement region is located at 5.5\arcsec\ north of the nucleus, co-spatial with the blueshifted emission observed in the \oiii velocity map and reaching values of $\sigma\sim$300~\kms in \oiii.
Interestingly, the southern cone exhibits a donut-shaped structure with a diameter of $\sim$0.6~kpc in all emission lines. A similar morphology, with a smaller size of $\sim$0.3~kpc, was reported in the nuclear region of NGC~1365 in [\ion{Ne}{V}]$\lambda14\,\mu$m by \cite{Ceci2026_1365}, where it was associated with the approaching side of the outflow. Similar bubble-like structures have also been observed in several AGN in ionized gas emission, where they are often interpreted as signatures of outflows and/or jet-ISM interactions, although on different spatial scales and not always accompanied by enhanced velocity dispersion \citep[e.g., ][]{May&Steiner2017,Maksym2017,May2020,Venturi2023}. 
However, in \ngc, the structure is not aligned with the ionization cone or the jet axes in the SW direction, but is instead oriented toward the south (see top right panel in Fig.~\ref{fig: moka results}). 
This suggests that outflowing gas clouds within the ionization cone follow slightly irregular trajectories rather than a smooth conical pattern, as expected when an outflow or jet interacts with a dense, inhomogeneous medium.
The velocity dispersion in this region allows a direct comparison with the near-IR results of \cite{Greene2014}, as it is characterized by low $A_V$ values (see Fig.~\ref{fig: attenuation map}). Within the SINFONI FoV, they traced the upper portion of the donut-shaped structure that we detect, although they reported lower $\sigma$ values (150$-$200~\kms). As discussed above, this difference is likely due to our higher sensitivity to diffuse gas.

The most intriguing feature is the $\sigma$-enhancement observed in the galaxy disk plane, perpendicular to the jet and outflow direction (see Fig.~\ref{fig: rgb image}). Present in all the second-moment emission line maps with values of 150$-$200~\kms, it is most prominent in the \oiii map, where it reaches the edge of the FoV ($\sim$2~kpc). We discuss about its origin in the following section.

\subsubsection{Physical origin of the $\sigma$-enhancement}
The $\sigma$-enhancement observed in the ionized gas velocity dispersion maps (see Fig.~\ref{fig: moment maps}) may result from the interaction between the radio jet and the host galaxy ISM, through the injection of perturbations in the direction perpendicular to the jet \citep[see e.g.][]{Querejeta2016,Girdhar2022,Audibert2023,PeraltadeArriba2023}. As discussed in \cite{Venturi2021_turmoil} and \cite{Ulivi2024}, an increased velocity dispersion over an extended region perpendicular to the jet and ionization cone axis is observed exclusively in AGN hosting small-scale ($\sim$1~kpc), low-power ($\leq 10^{44}$~erg~s$^{-1}$) radio jets with low inclinations ($\leq 45^\circ$) with respect to the galaxy disk. 
Alternative mechanisms, such as perturbations driven by SF or supernova feedback, are unlikely to account for the observed values of 150$-$200~\kms, since SF-driven perturbations in local galaxies typically yields velocity dispersions of $\lesssim$80~\kms \citep{Zhou2017,OlivaAltamirano2018}. 
In addition, bar-induced velocity dispersion enhancements are typically observed in the stellar component as a consequence of the orbital structure of the bar (e.g. \citealt{Jin2025}), and are not expected to produce the extended, gas-phase $\sigma$-enhancement with the geometry observed here.
We therefore favor a jet-driven origin for this feature.
For the first time, we observe this phenomenon in a galaxy hosting a radio jet perpendicular to the galaxy disk. As shown by the \oiii W$_{70}$\footnote{W$_{70}$ = v$_{85}$--v$_{15}$, where v$_{85}$ and v$_{15}$ are the 85th and 15th percentile velocities of the fitted line profile.} map in Fig.~\ref{fig: w70}, the $\sigma$-enhancement regions reach values of 300$-$400~\kms. In objects with radio jets lying close to the plane of the disk, these values can reach up to 1000~\kms \citep{Venturi2021_turmoil, Ulivi2024}.
This inclination-based difference is further supported by recent hydrodynamic simulations of jet-ISM interaction, which show that when the inclination angle of the radio jet with respect to the galactic disk is large, the jet can escape relatively easily, experiencing limited interaction with the disk material and producing less broadened emission-line profiles. Conversely, when coplanar with the disk, or at low inclination angles with respect to it, the jet remains trapped within the disk, strongly and extensively perturbing the ISM, resulting in a widespread increase of velocity dispersion \citep{Mukherjee2016, Mukherjee2018a, Mukherjee2018b, Meenakshi2022}.
The observed $\sigma$-enhancement could in principle be attributed to the superposition of multiple projected velocities from clouds in the rotating disk integrated along the line of sight (LOS), especially in a highly inclined system such as \ngc. However, the following argument disfavor a simple uniform projection scenario. As shown by the contours in Fig. \ref{fig: w70}, the regions of strongest attenuation spatially coincide with the $\sigma$-enhancement within the disk. This indicates that we are detecting the broad emission component in regions where the narrow disk emission is significantly obscured, whereas in less attenuated areas the narrow component dominates (see Fig. \ref{fig: rgb image}). If the $\sigma$-enhancement were purely a projection effect, it would be expected to occur more uniformly across the disk. Instead, its spatial correlation with dust attenuation suggests that the effect of jet-ISM interaction (i.e., the $\sigma$-enhancement) is likely present throughout much of the galaxy disk, but becomes detectable primarily in regions where dust attenuation suppresses the narrow emission component.
Further evidence supporting a jet–ISM interaction is provided by the morphology of the southern jet. In contrast to its northern counterpart, the southern jet is shorter and displays a secondary emission peak (i.e. a knot, see Fig.~\ref{fig: rgb image}). In addition, an enhancement in the \oiii flux is observed along the SW edge of the jet (see Fig.~\ref{fig: moment maps}). These features are indicative of jet-ISM interaction with possible shock-heating of the ISM due to the jet penetrating denser region of ISM.
The geometry of the jet and outflow axes relative to the nuclear region structures is visually summarized in the inset panel of Fig.~\ref{fig: attenuation map}, which schematically shows the relative orientation of the jet, the large-scale disk, the bar, and the nuclear disk.

The spatial similarity between the ionized gas $\sigma$-enhancement and the stellar velocity dispersion distribution complicates the interpretation. In bottom-right panel of Fig.~\ref{fig: moment maps} we overlay the \oiii second-moment contours on the stellar velocity dispersion map. The stellar velocity dispersion is overall lower ($\lesssim 110$~\kms) and more uniformly distributed across the disk than that of the gas, and broadly consistent with regular rotation (see Sect.~\ref{sec: stellar kinematics}). Nevertheless, its spatial distribution roughly traces the same regions as the \oiii\ second-moment map, complicating a straightforward interpretation of the $\sigma$-enhancement as purely jet-driven turbulence. 
We also inspected the CO(2--1) emission from ALMA in Appendix~\ref{appendix: CO moment maps, PV diagram}. The second-moment map shows only a marginal, spatially limited correspondence with the $\sigma$-enhancement toward the east, without conclusive evidence for a genuine molecular counterpart. From the position-velocity diagram (see Fig.~\ref{fig: PV diagram}) we identify two distinct kinematic components in the molecular gas: one associated with the large-scale rotating disk and another with the bar. The stellar component is dominated by the large-scale rotating disk, while in the nuclear region the stars trace the non-circular motions driven by the bar (e.g., \citealp{Greene2014}).
We discuss the implications of this in Sect.~\ref{sec: nature sigma}.

\subsection{Attenuation and diagnostic diagrams}
\label{sec: attenuation and BPT}
\begin{figure}
    \centering
    \includegraphics[width=0.99\linewidth]{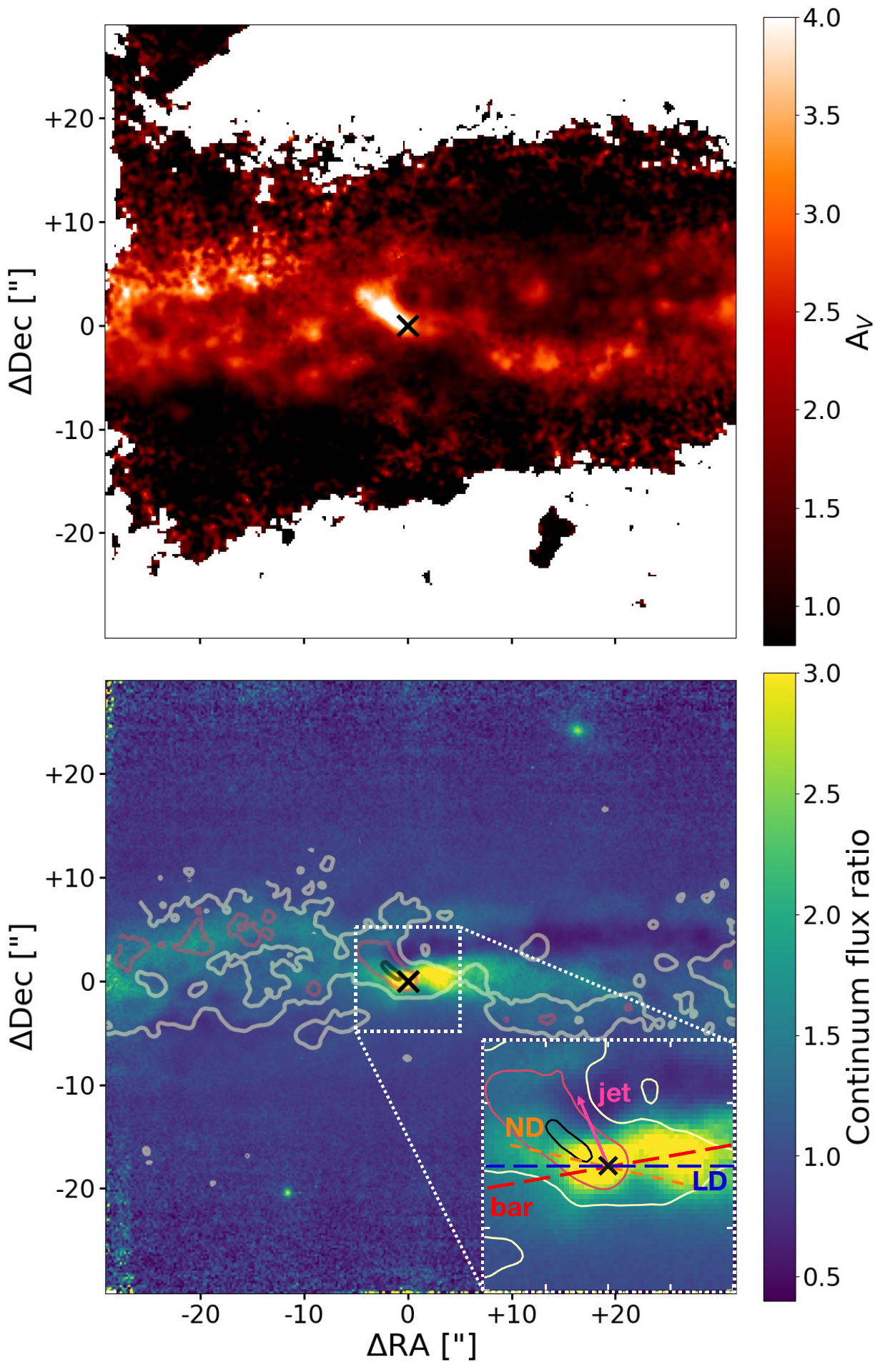}

    \caption{Dust attenuation and continuum color maps of \ngc. \textit{Upper}: $V$-band attenuation map obtained from the Balmer decrement and the \cite{Calzetti+2000} attenuation law. Spaxels with S/N $\leq$ 3 are masked. \textit{Lower}: map of the ratio between the mean continuum flux densities in the red ($\sim$9110--9120~\AA) and blue ($\sim$4780--4850~\AA) spectral intervals.
    The inset panel shows a zoom-in of the nuclear region ($5\arcsec \times 5\arcsec$, i.e. $\sim 0.4 \times 0.4$~kpc$^2$). We schematically indicate the jet orientation (magenta arrow; see Fig.~\ref{fig: rgb image}), the bar at PA = $100^{\circ}$ (red dashed line; \citealp{Veilleux1999a}), the large-scale disk at PA = $90^{\circ}$ (LD; blue dashed line), and the nuclear disk at PA = $75^{\circ}$ (ND; orange dashed line; \citealp{Greene2014}). 
    Contour levels correspond to $A_V = 2$, 3, and 4.5, shown in white, red, and black, respectively. The cross marks the position of the nucleus from the VLA 6~GHz image.}
    \label{fig: attenuation map}
\end{figure}

\begin{figure*}
    \centering
    \includegraphics[width=.8\linewidth, trim=10 10 9 10, clip]{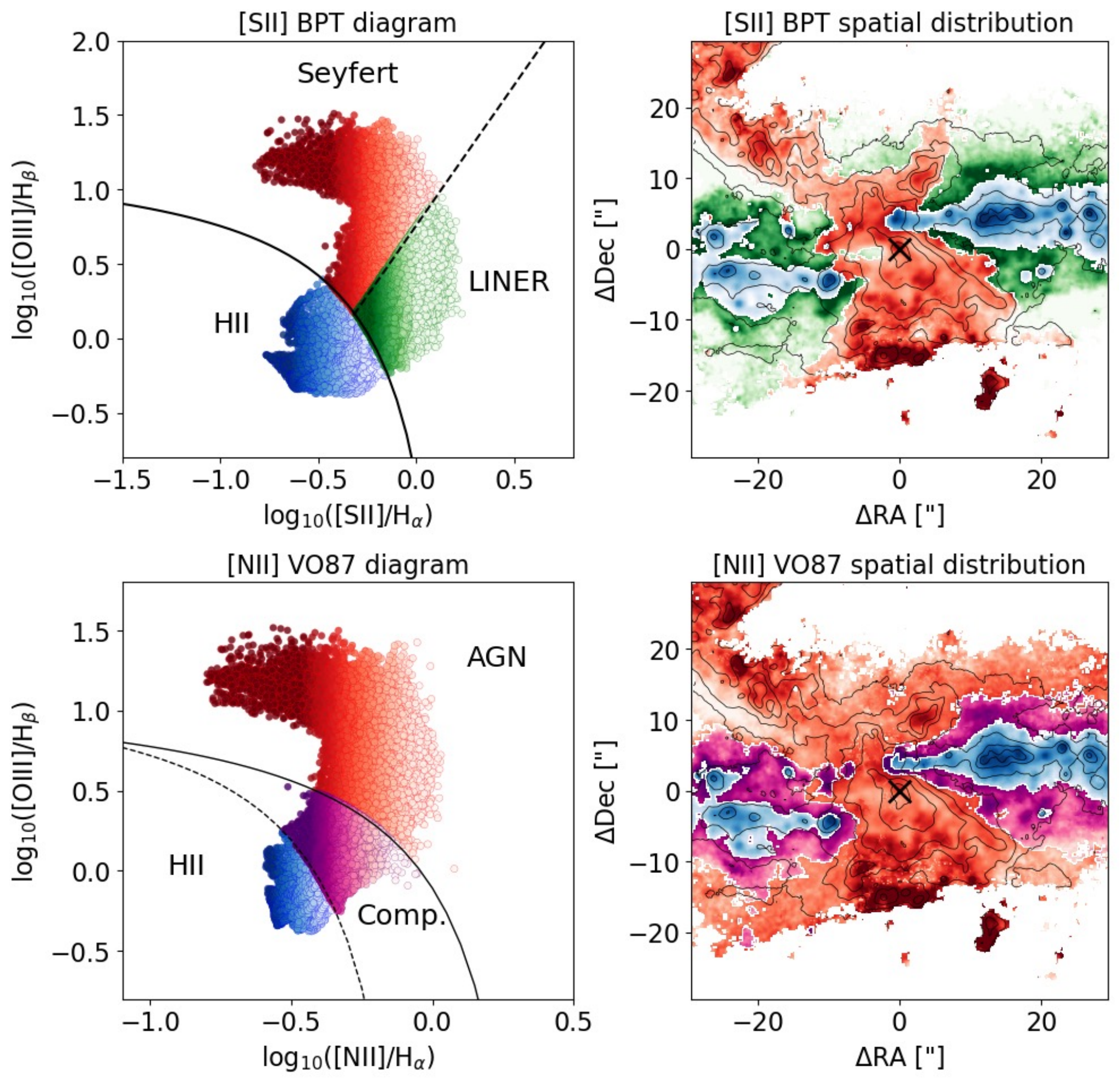}
    \caption{Spatially resolved BPT diagrams (left column) of \ngc\ and corresponding spatial distribution (right column).
    \textit{Upper panels}: [\ion{S}{II}]-BPT diagram of \siiALL/\halpha versus \oiii/\hbeta. 
    \textit{Lower panels}: [\ion{N}{II}]-VO87 diagram of \nii/\halpha versus \oiii/\hbeta. 
    The solid curve marks the theoretical upper limit for pure SF \citep{Kewley+01} in both diagrams. 
    In the BPT diagram, the dashed curve separates Seyfert galaxies from LINERs \citep{Kewley+06}, while in the VO87 diagram it corresponds to the empirical classification by \citet{Kauffmann2003}, with all bins below this curve being dominated by SF.
    Blue denotes SF-dominated regions, purple (green) intermediate regions in the VO87 (LINER-like in the BPT), and red AGN-ionized spectra, where the color lightness increases linearly with $x$-axis line ratios.
    Black contours represent arbitrary \oiii flux levels.
    The cross marks the position of the nucleus from the VLA 6-GHz image.
    Spaxels with S/N < 2 are masked.
    }
    \label{fig: bpt diagrams}
\end{figure*}

We compute the dust attenuation using the Balmer decrement (i.e., H$\alpha$/H$\beta$), assuming a case-B hydrogen recombination value of 2.86 for this ratio \citep[for an electron temperature and density of $T_e=10^4$~K and $n_e=10^2$~cm$^{-3}$, respectively;][]{Osterbrock2006} and adopting the \cite{Calzetti+2000} attenuation law. The $V$-band attenuation ($A_V$) map is shown in Fig. \ref{fig: attenuation map} (upper panel).
$A_V$ ranges from $\sim$1 to $\sim$5 within the disk, with the highest values found in a $\sim$6\arcsec\ ($\sim$0.5~kpc) structure extending to the NE of the nucleus. 
Clumpy regions with relatively high $A_V$ are also present across the disk, particularly along the dust lanes (see also Fig.~\ref{fig: rgb image}). 
In the northern outflow cone, the S/N is too low to draw conclusions, but in the lower cone and in the northeastern plume, the attenuation is clearly lower than in the disk ($A_V<$ 1). This supports the findings of \citet{Mingozzi2019} and \citet{ Ceci2026_1365}, which show that outflows in local Seyfert galaxies are less attenuated by dust than the host galaxy disk. Such a configuration is expected if the approaching outflow component, lying in front of the disk, dominates the optical emission, while the receding component remains obscured behind the disk and becomes visible only at larger distances, where the attenuation is reduced. This interpretation is consistent with the prominent detection of the southern outflow compared to the northern one in \ngc.
We note, however, that the relative attenuation between outflowing gas and host-galaxy ISM is not universal. Depending on the outflow geometry and dust distribution, the opposite configuration is also possible: for example, in the interacting system Mrk~848, \citet{Perna2019} found that the two nuclei display contrasting behaviors, with one showing less-attenuated outflows and the other more-attenuated outflows than the systemic emission.

In the lower panel of Fig.~\ref{fig: attenuation map}, we overplot the attenuation levels on a color image showing the ratio between the mean continuum flux densities measured in the red ($\sim$9110--9120~\AA) and blue ($\sim$4780--4850~\AA) spectral ranges.
Both the SF regions along the spiral arms (traced by the continuum emission; see Fig.~\ref{fig: rgb image}) and a nuclear structure are clearly visible. 
Interestingly, the latter appears asymmetric and elongated toward the west. Considering the inclinations of the bar and the nuclear disk reported by \cite{Veilleux1999a} and \cite{Greene2014} (see the inset panel in Fig.~\ref{fig: attenuation map}), this morphology may arise from the projected overlap of these two disk components, with the northern side corresponding to the near side of the galactic disk.
The elongated high-$A_V$ structure is aligned with the NLR traced by the [\ion{Si}{VI}] and Br$\gamma$ emission (\citealp{Greene2014}) and does not appear to show any clear correlation with the continuum color map. A possible interpretation is that the northern ionization cone illuminates gas in the galactic disk, and the resulting emission is strongly attenuated by dust in the host galaxy. By contrast, the southern ionization cone lies in front of the galactic plane from our point of view and therefore suffers much less attenuation.

We consider the classical emission-line diagnostic diagrams for gas excitation (\citealp{Baldwin1981,Veilleux1987}), based on $\log_{10}([\ion{O}{III}]\lambda5007/\mathrm{H}\beta)$ versus $\log_{10}([\ion{S}{II}]\lambda\lambda6717,6731/\mathrm{H}\alpha)$ and $\log_{10}([\ion{O}{III}]\lambda5007/\mathrm{H}\beta)$ versus $\log_{10}([\ion{N}{II}]\lambda6583/\mathrm{H}\alpha)$, hereafter referred to as the BPT and VO87 diagrams, respectively.
In the upper (lower) panel of Fig.~\ref{fig: bpt diagrams}, we show the spatially resolved BPT (VO87) diagram and the corresponding spatial distribution of the classified spaxels. 
As shown in the right-hand panels of Fig.~\ref{fig: bpt diagrams}, both the BPT and the VO87 reveal a clear separation between SF regions, intermediate and LINER or shock-like excitation, and AGN-dominated ionization. 
The majority of the SF-dominated spaxels lie along the host galaxy disk and trace the spiral arms, consistent with the morphology observed in Figs.~\ref{fig: rgb image} and \ref{fig: moment maps}. Specifically, the spaxels with lowest line ratios (displayed in dark blue) trace clumpy regions, likely associated with young stellar clusters \citep[e.g., ][]{Xiao2018}. 
LINER or shock-like regions (green in the VO87) are mainly located in the inner disk and around the edges of the outflow structure, suggesting a contribution from shocks produced by the interaction between outflow and the ISM. 
AGN-like ionization (red) dominates along the biconical outflow, extending several tens of arcseconds from the nucleus in both directions, consistent with the morphology traced by \oiii (Figs.~\ref{fig: rgb image} and \ref{fig: moment maps}). Interestingly, the spaxels with the higher ionization continuum (dark red) are located in the innermost regions of the outflow. This feature was also observed by \cite{Mingozzi2019} in the galaxies of the MAGNUM sample, suggesting that the innermost regions of the cone are possibly directly heated by the central ionizing AGN, being characterized by high excitation. 

To further investigate the connection between gas excitation and kinematics, in Fig.~\ref{fig: bpt sigma} we show the same BPT and VO87 diagrams color-coded by the \oiii velocity dispersion. Most of the spaxels with higher $\sigma$ are AGN-ionized, in agreement with the results of \cite{Ulivi2024}. Moreover, the $\sigma$-enhanced regions perpendicular to the radio jet (i.e., within the galaxy disk) exhibit LINER-like ionization, often associated with shocked material, as also observed by \cite{Venturi2021_turmoil} in other local Seyferts with low jet-disk inclination angles. The scenario of an interaction between the radio jet and the ISM is further supported by the fact that the elevated [\ion{S}{II}]/\halpha and [\ion{N}{II}]/\halpha ratios, observed in regions of high $\sigma$, are well reproduced by shock models \citep{Allen2008}, with shock velocities in the range 100--1000~\kms (see details in \citealt{Mingozzi2019}).

\subsection{Outflow kinematics, energetics, and jet-ISM interaction} \label{sec: energetics}
In the following, we characterize the energetics of \ngc\ through two complementary approaches. First, we model the biconical outflow using a 3D kinematic model and compare its radial velocity profile with predictions from an energy-conserving AGN wind model (Sects.~\ref{sec: moka fit results}-\ref{sec: vel profile model}). Second, we assess the role of the radio jet in driving the enhanced velocity dispersion observed in the galaxy disk (Sect.~\ref{sec: comparison with literature}). Although these two mechanisms are analyzed independently, their interplay is discussed in Sect.~\ref{sec: nature sigma}.

\subsubsection{\moka kinematic modeling} \label{sec: moka fit results}
We inferred the intrinsic (i.e., de-projected) radial velocities of the outflow ($v_{\rm out}$) using the 3D multi-cloud \moka\footnote{\href{https://github.com/cosimomarconcini/moka_3d}{https://github.com/cosimomarconcini/moka$\_3$d}.} kinematic model \citep{Marconcini2023}. This model has demonstrated its capability to robustly reproduce extremely detailed outflow features, exploiting IFU data from a variety of instruments, including VLT/MUSE \citep{Marconcini2023, Marconcini2025_nat}, JWST/NIRSpec \citep{Cresci2023,Perna2024, Ulivi2025} and JWST/MIRI \citep{MIRACLE_NGC424, Marconcini2025, Ceci2026_1365}. 
Given the projected overlap between the outflow and the galaxy disk, we followed the approach of \cite{Marconcini2025_nat} modeling the observed \oiii\ line profile as the combination of two kinematic components, associated with a biconical outflow and a thin rotating disk, respectively. For the outflow component, the cones were fitted independently from each other. We chose \oiii\ because it is the warm-ionised transition with the highest S/N in the outflow component and is not blended with other strong optical lines. Full details of the biconical outflow and thin-disk modelling are given in Appendix~\ref{app: moka modelling setup}.
The best-fit values for the disk and outflow parameters are listed in Table~\ref{tab:moka_fit_outflow}.

\begin{figure*}
    \centering
    \includegraphics[width=.99\linewidth, trim=18 10 14 10, clip]{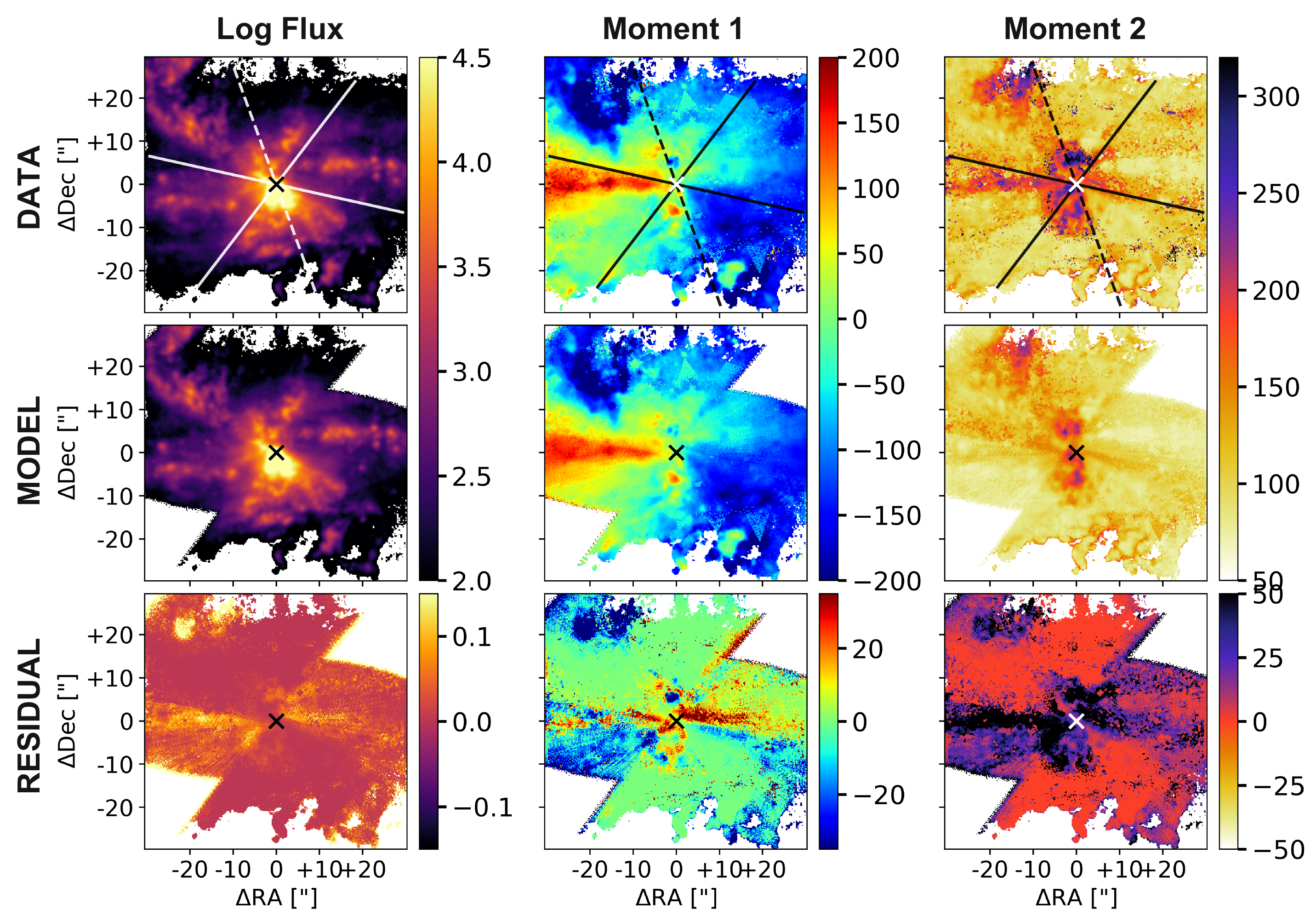}
\caption{Comparison between observed and modeled moment maps for \ngc. From top to bottom, panels show the moment maps from the \oiii emission-line fitting (same as in Fig.~\ref{fig: moment maps}), those resulting from \moka modeling, and the residuals between them. From left to right: integrated flux (in logarithm scale, in units of 10$^{-17}$~erg~s$^{-1}$~cm$^{-2}$), LOS velocity and velocity dispersion maps (both in \kms) from \oiii emission. The dashed line marks the outflow axis (PA~=~$20^\circ$), while the solid lines indicate the outflow boundaries for an opening angle of $115^\circ$, as derived from the best-fitting \moka model (see Sect.~\ref{sec: moka fit results} and Table~\ref{tab:moka_fit_outflow}). The cross marks the position of the nucleus from the VLA 6~GHz image.}
    \label{fig: moka results}
\end{figure*}

In Fig. \ref{fig: moka results} we compare the observed and modeled moment maps obtained with \moka. The residual maps (bottom row) indicate that the fit performs better in the outflow region, except at larger distances where the S/N decreases or where residual structures arise from imperfect modeling of the three-dimensional velocity field or geometry of the outflowing cones (e.g., the NE plume). The larger residuals observed in the disk region are likely due to non-rotational motions or deviations from an idealized rotating disk structure (such as bulges and bars, or due to the presence of the jet; \citealt{Haidar2026,Rosario2026}), particularly in the area where we detect the $\sigma$ enhancement (see Fig. \ref{fig: w70}). 
In any case, these features do not influence our modeling of the outflow properties. It is important to emphasize that the contribution from the underlying disk emission is negligible when modeling and reproducing the intrinsic properties of the outflow, given the considerable differences in projected velocity between outflow and disk. Indeed, from the disk fit we found a rotation velocity of 190 \kms, significantly lower than the intrinsic outflow velocities ($\sim 400$~\kms), and therefore weakly affecting our results.

The radial averaged properties of the outflow and disk kinematics and geometry are summarized in Table~\ref{tab:moka_fit_outflow}. We stress that the inclination angle of the outflow ($\alpha$) and disk ($\beta$) obtained are basically the same ($\sim75^{\circ}$), indicating that the outflow axis is mostly perpendicular to the disk plane. We also stress that the outflow velocity ($v_{\rm out}$) inferred by \moka\ corresponds to the de-projected, intrinsic outflow velocity.
As a result, we also obtained the outflow velocity as function of distance from the nucleus (see Fig.~\ref{fig: vel profile model}). 
We discuss the radial velocity pattern of the outflow and apply a physically motivated, energy-conserving model for an AGN-driven outflow in Sect.~\ref{sec: vel profile model}.

\begin{table}[t]
\centering
    \caption{Best-fit parameters for the outflow and disk properties in \ngc, inferred within the \moka\ framework (see also Sec.~\ref{sec: energetics}).}
    \label{tab:moka_fit_outflow}
    \setlength{\tabcolsep}{3pt}
    \begin{tabular}{llcc}
   \hline
   \hline
    Parameter                  & Free &  Fit \\
    \hline
    Outflow &  &\\
   \hline
    $\alpha$ ($^{\circ}$)    & Y             & 76 $\pm$ 3 (Northern)  \\
                             &               & 88.0 $\pm$ 1.1 (Southern)   \\
    $v_{\rm out}$ (\kms)     & Y             & 430 $\pm$ 6 (Northern) \\
                             &               & 380 $\pm$ 60 (Southern)   \\
   PA ($^{\circ}$)           & N             &  20  \\
   $\theta$ ($^{\circ}$)     & N             &  115  \\
   $R_{\rm max}$             & N             & 38\arcsec (3.1 kpc) \\
    \hline 
   Disk &  &\\
    \hline 

   $\beta$ ($^{\circ}$)     & Y            & 75 $\pm$ 3   \\
   $v_{\rm rot}$ (\kms)         & Y            & 190 $\pm$ 50 \\
   PA ($^{\circ}$)              & N            &  90    \\
   $R_{\rm max}$                & N &          40\arcsec (3.2 kpc) \\
   \hline 
  \end{tabular}
\tablefoot{From left to right: Parameter, flag indicating whether the parameter was free to vary (Y) or fixed (N) in the fitting, and the radial averaged best-fit value. The fitted parameters are the outflow inclination ($\alpha$; angle between the cone axis and the the LOS), the mean outflow intrinsic radial velocity ($v_{\rm out}$), the disk inclination angle ($\beta$; relative to the LOS), and the disk circular velocity ($v_{\rm rot}$). The outflow and disk PAs are kept fixed during the fit, as well the opening angle $\theta$. $R_{\rm max}$ is the maximum radius of the disk and the outflow, set by the limited FoV of the MUSE data. For details on the conical outflow structure and parameters,  see Sec. \ref{sec: energetics}.}
\end{table}


From the best-fit model obtained with \moka, we computed the energetics of the two outflow cones with \oiii. 
We first estimated the electron density of the ionized gas from the \siiALL line ratio \citep{Osterbrock2006}, assuming an electron temperature of $T_{\rm e} = 10^4$ K. By separately integrating the flux of the two outflow cones over the same mask adopted for the \moka\ modeling, we derived electron densities of $(190 \pm 110)\,\mathrm{cm}^{-3}$ and $(400 \pm 80)\,\mathrm{cm}^{-3}$ for the northern and southern cones, respectively. To account for the uncertainty associated with the electron temperature, we performed a Markov Chain Monte Carlo analysis \citep{emcee} using a uniform prior on $T_{\rm e}$ in the range $10^{3-5}$ K.
We found that the average electron density variations remain within the uncertainties of the value inferred assuming $T_{\rm e}$ = 10$^4$\,K.

We computed the outflow mass for each cone following \cite{Ceci2026_1365} (see their Appendix G), using the \oiii flux extracted from the best-fit \moka model (corrected for dust attenuation) and the electron densities derived above. We assumed solar oxygen abundance $12 + \log(\mathrm{O/H})_\odot = 8.69$ \citep{Asplund2009}.
Then, we estimated the ionized outflow energetics, that is, its mass outflow rate ($\dot{M}_{\rm out}= M_{\rm out} v_{\rm out} / R_{max}$), kinetic energy ($ E_{\rm out} = \frac{1}{2} M_{\rm out} v_{\rm out}^2$), kinetic power ($ \dot{E}_{\rm out} = \frac{1}{2} \dot{M}_{\rm out} v_{\rm out}^2$), and momentum rate ($ \dot{P}_{\mathrm{out}} = \dot{M}_{\rm out} v_{\rm out}$), adopting the intrinsic outflow velocity derived with \moka\ separately for the northern and southern cones (see Table~\ref{tab:moka_fit_outflow}).
Table~\ref{tab: outflow_energetics} shows the comparison of the aforementioned outflow properties for the two cones and for the total outflow (obtained as the sum of the quantities from each cone).
\cite{Lamperti2020} found a star formation rate of SFR~$\sim0.8$~M$_{\odot}$~yr$^{-1}$ and a molecular gas mass of $M_\text{mol}\sim8\times10^7$~M$_{\odot}$ in \ngc, noting that the two quantities were derived from different apertures (IR SED fitting for the SFR, CO(3--2) with beam correction for $M_\text{mol}$). 
Considering the mass outflow rate for the total outflow, we compute a mass loading factor of ($\dot{M}_\text{out}$/SFR)~$\approx0.28$ and an outflow depletion time of ($M_\text{mol}$/$\dot{M}_\text{out}$)~$\approx350$~Myr. The mass-loading factor below one suggests that the ionized outflow alone is unlikely to quench SF, although the total value may be higher when including molecular and neutral atomic phases \citep{Fluetsch2019, Rupke2021}. The relatively low $M_\text{mol}$ could also be affected by ongoing \ion{H}{I} stripping in the Virgo cluster \citep{Oosterloo2005, Vollmer2018} and by the aperture mismatch with the SFR estimate.
Despite these caveats, the depletion time is significantly shorter than the $\sim2$~Gyr typical of star-forming galaxies \citep{Leroy2013}, indicating that the outflow may be contributing to gas depletion.

\begin{table}[t]
    \centering
    \caption{Ionized outflow properties in \ngc\ traced by \oiii emission line.}
    \label{tab: outflow_energetics}
    \begin{tabular}{lccc}

        \hline
         & Northern & Southern & Total \\
        \hline

        $M_{\rm out}$ (10$^{5}$ \msun) 
            & 7 $\pm$ 4     
            & 4.6 $\pm$ 1.0     
            & 11 $\pm$ 5 \\

        $\dot{M}_{\rm out}$ (10$^{-1}$ \msun yr$^{-1}$)  
            & 1.3 $\pm$ 0.7 
            & 1.0 $\pm$ 0.2  
            & 2.3 $\pm$ 1.0 \\

        $E_{\mathrm{\rm out}}$ (10$^{54}$ erg) 
            & 2.2 $\pm$ 1.3  
            & 2.1 $\pm$ 0.5 
            & 4.3 $\pm$ 1.8 \\

        $\dot{E}_{\mathrm{out}}$ (10$^{40}$ erg s$^{-1}$) 
            & 1.3 $\pm$ 0.8
            & 1.5 $\pm$ 0.4
            & 2.8 $\pm$ 1.2 \\

        $\dot{P}_{\mathrm{out}}$ (10$^{32}$ dyne) 
            & 5 $\pm$ 3
            & 4.4 $\pm$ 1.1
            & 9 $\pm$ 4 \\

\hline
    \end{tabular}
    \tablefoot{
    From top to bottom: outflow mass ($M_{\rm out}$), mass outflow rate ($\dot{M}_{\rm out}$), kinetic energy ($E_{\mathrm{out}}$), kinetic power ($\dot{E}_{\mathrm{out}}$), and momentum rate ($\dot{P}_{\mathrm{out}}$). The values are derived by extracting integrated quantities over the FoV from the \moka\ outflow modeling results (see Sect. \ref{sec: energetics} for details). For comparison, we list the estimates for the Northern and Southern outflow cones, as well as for the total biconical outflow\footnote{Differences between the total column and the sum of the other columns are attributable to rounding.}.}
\end{table}

\subsubsection{Comparison of outflow-driving mechanisms} \label{sec: comparison outflow mechanisms}
Localized jet-ISM interactions may occur in the innermost regions of a galaxy \citep[e.g.,][]{Mukherjee2016, Tanner2022}, so the low-power radio jet may be the primary driver of the kpc-scale ionized outflow in \ngc. The jet kinetic power, $P_{\rm jet} \sim 10^{42}$~erg~s$^{-1}$ (see  Sect.~\ref{sec: comparison with literature}), exceeds the outflow kinetic power ($\dot{E}_{\rm out} \simeq 2.8 \times 10^{40}$~erg~s$^{-1}$; see Sect.~\ref{sec: moka fit results}) by about a factor of 30.
Jet-driven outflows typically require a jet power one to two decades larger than the outflow kinetic power \citep{Mukherjee2016, Tanner2022}, so the jet could in principle be the driver of the observed biconical outflow.

On the other hand, adopting the hard X-ray anchored bolometric luminosity $L_{\rm bol} \simeq 9 \times 10^{43}$~erg~s$^{-1}$ \citep{Koss22}, the power of the AGN radiation-driven nuclear wind expected in the \citet{King2015} framework is $P_{\rm wind} \simeq 0.05 \times L_{\rm bol} \simeq 4 \times 10^{42}$~erg~s$^{-1}$. In the energy-conserving scenario, this wind power is expected to be almost entirely transferred to the large-scale outflow, i.e. $\dot{E}_{\rm out} \sim P_{\rm wind}$, implying $\dot{E}_{\rm out}/L_{\rm bol} \sim 5\%$ \citep{Zubovas&King2012}. The observed coupling efficiency ($\dot{E}_{\rm out}/L_{\rm bol} \sim 0.03\%$) is more than two orders of magnitude lower than this idealized prediction, and also below the more realistic estimate of $\dot{E}_{\rm out}/L_{\rm bol} \sim 0.5\%$ from \citet{Costa2014}. This suggests that the accretion-disk wind alone may not efficiently power the observed kpc-scale outflow, or that the coupling in \ngc\ is less efficient than typically assumed.

From the energetics point of view, since $P_{\rm jet} \sim P_{\rm wind}$, both mechanisms are capable of driving the outflow with a modest coupling efficiency of order $\sim$1\%. In the following section, we model the radial velocity profiles obtained with \moka\ (Sect.~\ref{sec: moka fit results}) neglecting the jet
contribution.

\subsubsection{Energy-conserving model for wind-driven outflow} \label{sec: vel profile model}
Although the jet and the AGN wind are energetically comparable (Sect.~\ref{sec: comparison outflow mechanisms}), we first adopt the energy-conserving wind framework of \citet{King2015}, which models the outflow as driven solely by the AGN accretion-disk wind (see Appendix~\ref{appendix: setup profile model} for details). We therefore neglect the jet contribution in the following analysis, acknowledging that jet-ISM interaction may affect the innermost kinematics \citep[e.g.,][]{Mukherjee2016, Tanner2022} and discuss it in detail in Sect.~\ref{sec: comparison with literature}.

\begin{figure}
    \centering
    \includegraphics[width=1.02\linewidth]{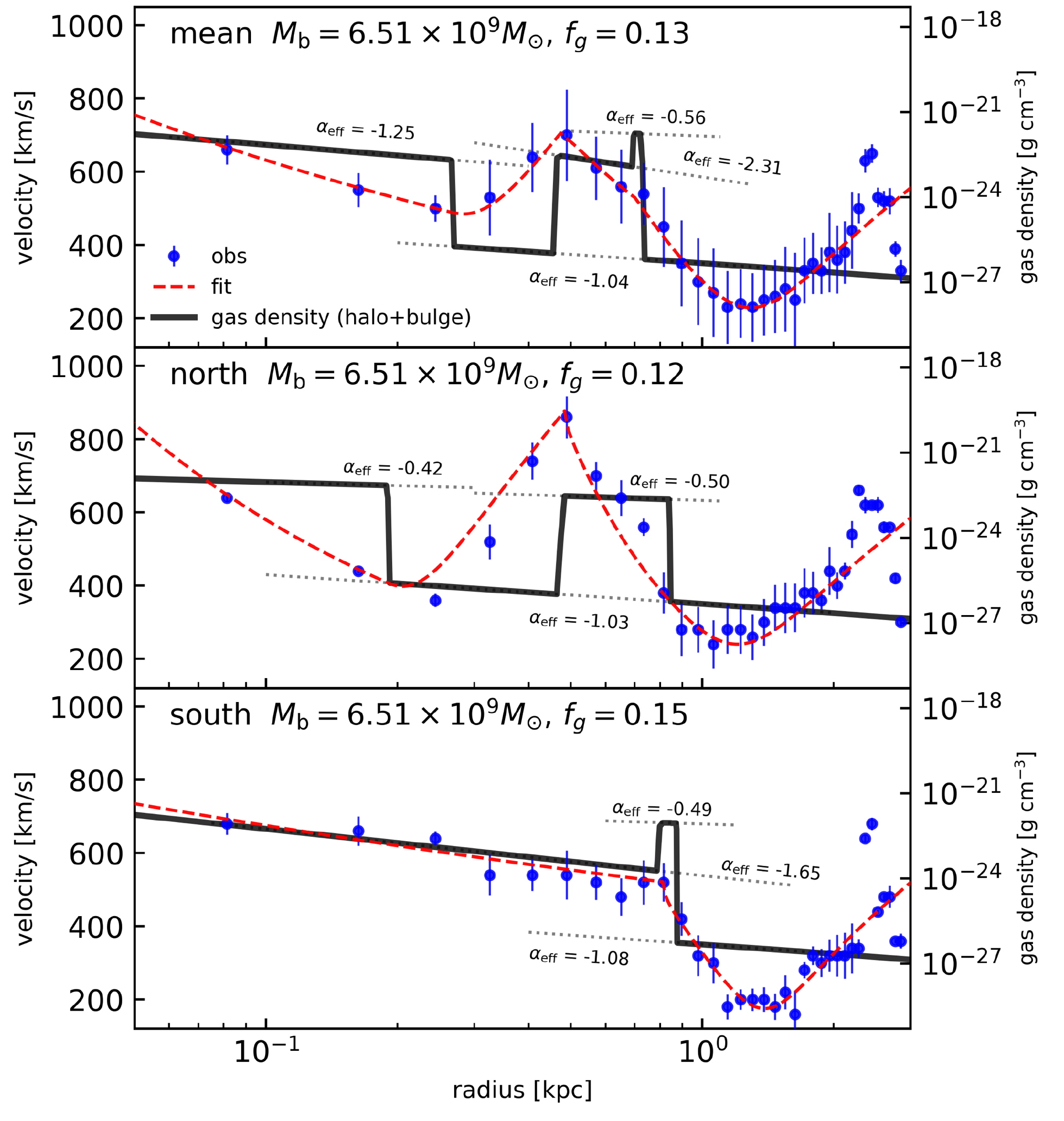}
    \caption{Outflow velocity profiles and best-fit models obtained with \textsc{MAGNOFIT} for the northern (middle), the southern (bottom) outflows and their mean (top). Blue points show the de-projected radial outflow velocities derived from the \oiii\ emission using \moka (see Sect.~\ref{sec: moka fit results}), while the red dashed lines represent the corresponding best-fit models. 
    The black solid lines indicate the underlying gas density profiles, with the right $y$-axis showing the corresponding density scale. The effective density slopes of each segment are also reported. 
    Best-fit values of the bulge mass ($M_\mathrm{b}$) and gas fraction ($f_\mathrm{b}$) are indicated in each panel.}
    \label{fig: vel profile model}
\end{figure}

To investigate whether the outflow in \ngc\ can be described by an energy-conserving outflow, we employed the 1D outflow evolution code MAGNOFIT, which assumes a multi-component bulge-halo system \citep{Zubovas2022}. This tool solves the equation of motion for an adiabatic outflow within a general spherically symmetric gravitational potential and gas density distribution. 
We applied the code to the two individual outflow velocity radial profiles of the north and south outflow cones, as well as to their average (Fig.~\ref{fig: vel profile model}), as derived with \moka in Sect.~\ref{sec: moka fit results}. The three fits were performed independently assuming a spherical symmetry, but sharing the same stellar velocity dispersion of the bulge and SMBH mass (more details in Appendix~\ref{appendix: setup profile model}).

As seen in Fig.~\ref{fig: vel profile model}, we find that a relatively narrow “obstruction” (i.e., a denser shell at $R \sim 0.8$ kpc) can account for the observed deceleration of the southern outflow. In contrast, reproducing the features of the northern outflow requires a more extended dense shell combined with a low-density gap. 
The good match between the outflow velocity radial profile predicted by MAGNOFIT and that inferred from the MUSE data through \moka supports an energy-conserving nature of the outflow, consistent with previous findings based on spatially resolved ionized-gas outflows in other nearby AGN \citep{Marconcini2025, Zubovas2025}.  We note, however, that this agreement relies partly on ad hoc density structures whose parameters are not independently constrained, so the fit should be regarded as consistent with the energy-conserving scenario rather than a definitive test of the energy-conserving hypothesis.
In this framework, the observed velocity field does not necessarily trace variations in the instantaneous AGN power, but rather the response of the expanding shocked bubble to the local density structure of the host galaxy. 
The different behaviors of the northern and southern outflows, therefore, suggest that the two cones are propagating through markedly different ISM conditions, with dense circumnuclear or extraplanar structures locally slowing down the flow and low-density channels allowing subsequent acceleration. 
This interpretation provides a simple explanation for the observed acceleration and deceleration patterns, although the inferred density structures should be regarded as effective one-dimensional representations of a more complex three-dimensional ISM.
Moreover, we cannot exclude the contribution of other driving mechanisms, such as direct AGN radiation pressure on dusty clouds \citep[e.g.,][]{Costa2018} or the radio jet. In either case, the resulting outflow kinematics would similarly be shaped by the inhomogeneity of the ISM along its propagation path \citep{Zakamska&Greene2014,Young2025}. 


\subsubsection{Radio jets as drivers of the $\sigma$-enhancement in the disk} \label{sec: comparison with literature}
Beyond the biconical outflow region, the velocity dispersion maps reveal an enhancement along the disk (Fig.~\ref{fig: moment maps}). By combining their results with those of \cite{Venturi2021_turmoil}, \cite{Ulivi2024} found evidence for a correlation between the kinetic power of the radio emission and the mass and energy of the turbulent ionized gas mostly located perpendicular to the jet. Their analysis indicates that the jets are sufficiently powerful to account for the observed kinetic energy in regions of enhanced line width, aiming at an efficient transfer energy to the surrounding medium. This suggests that the radio jets are responsible for driving the line broadening by interacting with the galaxy ISM and significantly perturbing it.

To assess whether this scenario is also applicable to \ngc, we estimated the mass and kinetic energy of the ionized gas associated with the $\sigma$-enhanced regions, and subsequently compared them with the jet power ($P_{\rm jet}$).
We first defined the $\sigma$-enhanced regions by selecting spaxels with residuals between the observed velocity-dispersion map and the best-fitting \moka\ model larger than $20$~\kms\ (bottom-right panel of Fig.~\ref{fig: moka results}). The above threshold ensures that we avoid the typical second-moment residual values in the galaxy disk region ($\lesssim$~10~\kms), where we do not observe any $\sigma$-enhancement (see, e.g., point~d in Fig.~\ref{fig: rgb image}). 

Following \cite{Venturi2021_turmoil} and \cite{Ulivi2024}, we computed the disturbed ionized gas using the total \halpha\ emission line profile (i.e., not limited to the broad component), integrated over the $\sigma$-enhanced regions defined above, including both outflowing and disk components, using the following relation:
\begin{equation}
    \frac{M_{\text{ion}}}{M_{\odot}} = 3.2 \times 10^{5} \left( \frac{L_{\mathrm{H}\alpha}}{10^{40} \, \mathrm{erg} \, \mathrm{s}^{-1}} \right) \left( \frac{n_{\mathrm{e}}}{100 \, \mathrm{cm}^{-3}} \right)^{-1} ,
    \label{eq: mass outflow}
\end{equation}
where $L_{\mathrm{H}\alpha}$ is the \halpha luminosity corrected for dust attenuation and $n_{\rm e}$ the electron density. 
We estimated the electron density from the \siiALL line ratio, measured from the total emission within the same masked region, yielding $n_{\rm e} = (300 \pm 70)\,\mathrm{cm}^{-3}$. 
We obtained $M_{\rm ion} = (1.3 \pm 0.3)\times 10^6\,M_{\odot}$. 
We also estimated the kinetic energy of the gas in the same region as $ E_{\rm ion} = M_{\rm ion}\sigma_{\rm ion}^2/2$, where $\sigma_{\rm ion}$ is the velocity dispersion of \halpha. 
We derived $E_{\rm ion} = (1.4 \pm 0.7)\times 10^{53}\,\mathrm{erg}$.
Leveraging the kinematic decoupling provided by \moka, we also estimated the energetics from the residual \oiii\ flux, which we interpret as tracing the turbulent gas actually affected by the jet-ISM interaction. We derived the outflow mass using Eq.~G.6 of \cite{Ceci2026_1365}, integrating the residual \oiii\ flux over the same regions adopted for the \halpha\ estimate (see Fig.~\ref{fig: moka results}). To account for the known discrepancy between [O\,\textsc{iii}]- and \halpha-derived masses, we applied a multiplicative factor of 3  \citep[e.g.][]{Carniani2015, Fiore17, Venturi2023}. We obtain $M_{\rm ion} = (8 \pm 2)\times 10^4~M_{\odot}$ and $E_{\rm ion} = (9 \pm 2)\times 10^{51}~\mathrm{erg}$.
We note that the two estimates above are not directly comparable. One relies on the total line flux, whereas the other isolates the residual emission from \moka. In addition, we stress out that these residuals may also reflect other coherent non-rotational motions not accounted for in the \moka\ fitting procedure. For instance, we did not include a bar component in the model, even though its presence is suggested by \citet{Veilleux1999a}, who identified a large-scale stellar bar from modeling the H$\alpha$ velocity field. We therefore regard the possible presence of a bar as a caveat in our interpretation, and our estimate of the gas mass perturbed by the jet-ISM interaction may be subject to systematic uncertainties and may be an upper limit.

\cite{Ulivi2024} estimated the jet kinetic power ($P_{\rm jet}$) from the radio luminosity at 1.4~GHz ($P_\text{1.4\,GHz}$) using the relation from \citet[][see also \citealt{Venturi2021_turmoil}]{Birzan2008}. In this work, we instead adopt the calibration from \cite{Cavagnolo2010}, which extends the \cite{Birzan2008} sample toward lower radio powers, better suited to our case. However, we caution that such empirical calibrations were derived for classical, evolved radio jets in the hot gaseous haloes of galaxies or clusters, and in particular \cite{Cavagnolo2010} was calibrated on giant elliptical galaxies; their direct application to a young jet propagating in the dense ISM of a Seyfert spiral galaxy such as \ngc\ may therefore introduce systematic uncertainties in the absolute value of $P_{\rm jet}$.
To derive $P_{1.4\,\mathrm{GHz}}$, we first computed the spectral index $\alpha$ from the  ratio between the VLA flux density images ($S_\nu \propto \nu^{\alpha}$) at 6 and 20~GHz. Assuming the same spectral index at lower frequencies, we then extrapolated the 6~GHz flux image to 1.4~GHz (see details in Appendix~\ref{appendix: Pjet}).
By spatially integrating the $P_{1.4\,\mathrm{GHz}}$ image and adopting the \cite{Cavagnolo2010} relation, we derived a jet kinetic power of $P_\text{jet}=1.4^{+2.0}_{-0.8}\times10^{42}$~erg~s$^{-1}$.
For comparison, adopting the \cite{Birzan2008} calibration yields $P_{\rm jet} \sim 9.4 \times 10^{42}$~erg~s$^{-1}$. This significant difference likely arises because the \cite{Birzan2008} sample does not adequately sample the low-$P_\text{1.4\,GHz}$ regime, which is instead better represented in the \cite{Cavagnolo2010} calibration.
\cite{Mukherjee2018b} found that simulation-based estimates of $P_{\rm jet}$ for IC~5063 exceed those from the \cite{Birzan2008} and \cite{Cavagnolo2010} relations by about an order of magnitude, although the simulation value remains consistent with the scatter of those relations. We therefore stress that our estimate of $P_{\rm jet}$ should be regarded as an order-of-magnitude indicator rather than a precise measurement.

We note that \cite{Zanchettin2023} measured $P_{\rm jet} \sim 6 \times 10^{42}$~erg~s$^{-1}$ in NGC~2992, a nearby Seyfert galaxy with properties broadly comparable to \ngc, including an almost edge-on disk and a well-defined biconical outflow.
The X-ray luminosities of the two systems from BASS are also similar within $\sim$0.3~dex ($\log$($L_\text{2-10~keV}/[\text{erg s$^{-1}$}]$) = 42.5 and 42.3 for NGC 4388 and NGC 2992, respectively; \citealp{Ricci17}).
\cite{Zanchettin2023} also reported $\sigma$-enhanced ionized gas ($\sim$250--300~\kms) in the nuclear region of NGC~2992, qualitatively similar to what we observe in \ngc. However, the radio emission in NGC~2992 is compact and the presence of a collimated jet remains uncertain \citep{Zanchettin2023}, which prevents a direct comparison of the jet--disk geometry between the two systems. Despite their similar jet powers, the difference in jet morphology and orientation supports a scenario in which the efficiency of jet-ISM coupling likely also depends on additional factors such as the jet geometry and the degree of confinement within the disk.

\begin{figure*}
    \centering
    \includegraphics[width=.99\linewidth, trim=14 10 21 10, clip]{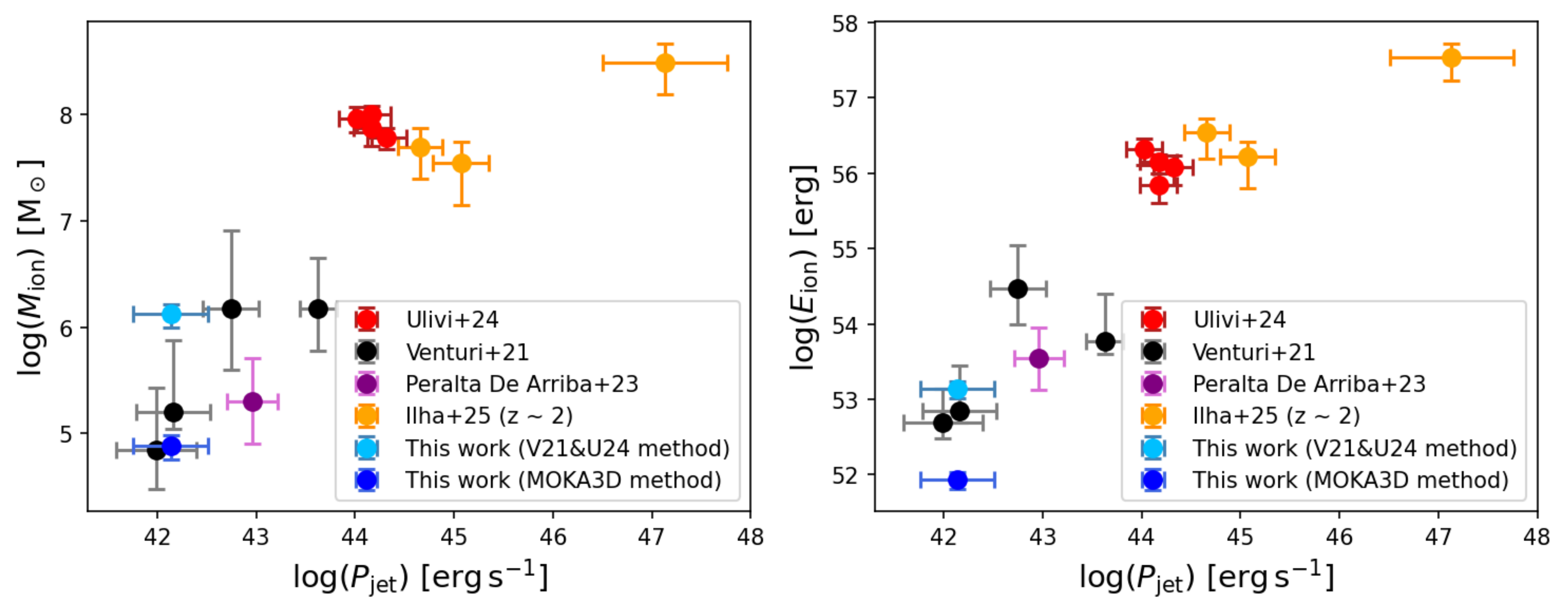}
    \caption{Ionized gas mass (left) and kinetic energy (right) as a function of jet kinetic power. Measurements for \ngc\ derived in this work are shown as cyan points when using the method of \cite{Venturi2021_turmoil} and \cite{Ulivi2024}, and as blue points when adopting the \moka\ method. Literature data are taken from \cite{Venturi2021_turmoil}, \cite{PeraltadeArriba2023}, \cite{Ulivi2024}, and \cite{Ilha2025}. For consistency, all jet powers have been recalculated using the \cite{Cavagnolo2010} calibration.}
    \label{fig: plot energetics}
\end{figure*}

In Fig.~\ref{fig: plot energetics}, we report the mass of turbulent gas and its kinetic energy as a function of the jet power for \ngc, compared to those obtained for local AGN-host galaxies by \cite{Venturi2021_turmoil}, \cite{ PeraltadeArriba2023} and \cite{Ulivi2024} ($P_\text{1.4\,GHz}\sim 10^{21}-10^{24}$ W Hz$^{-1}$; jet-disk inclination angles $\lesssim45^{\circ}$), and with results on three quasars at cosmic noon from \cite{Ilha2025}, spanning moderate to high radio powers ($P_\text{1.4\,GHz}\sim 10^{25}-10^{28}$ W Hz$^{-1}$). For consistency, we recomputed all literature $P_{\rm jet}$ values using the \cite{Cavagnolo2010} relation, as discussed above.
The ionized gas masses reported in \cite{PeraltadeArriba2023} and \cite{Ilha2025} were derived considering only the kinematic component identified as the outflow (based on velocity dispersion thresholds, separated from the systemic/disk component via multi-Gaussian line fitting), and are included as outflow-only reference points, whereas our estimates include all gas contributing to the line broadening in the $\sigma$-enhanced regions. Moreover, the masses of \cite{Ilha2025} are derived from [\ion{O}{III}], so we rescale them by the factor of 3 to account for their [\ion{O}{III}]-to-\halpha mass discrepancy, as explained above.
As noted by \cite{Ulivi2024}, both gas mass and kinetic energy correlate with jet power, extending over more than two orders of magnitude in jet power and $\sim$3 dex in both gas mass and kinetic energy. This trend suggests a tight correlation between the jet power and the mass and energetics of the ionized gas, supporting a scenario in which radio jets enhance the emission-line widths in directions perpendicular to their axes due to the interaction with the ISM. 
We quantify the observed trends using the Pearson correlation coefficient. Considering the full sample, we find a positive correlation between $P_{\rm jet}$ and $M_{\rm ion}$ ($r = 0.86$, $p = 8\times10^{-5}$), and a stronger correlation with the kinetic energy of the ionized gas ($r = 0.91$, $p = 7\times10^{-6}$).
We note, however, that this test does not account for measurement uncertainties or censored data.

In Sect.\ref{sec: comparison outflow mechanisms} we found that the jet could in principle be the driver of the observed biconical outflow from an energetic point of view. To assess whether the jet is sufficiently powerful to act as a viable mechanism of feedback, we compared the total kinetic energy released by the jet, $E_{\rm jet}$, over its propagation time, $t_{\rm jet}$, with the kinetic energy of the ionized gas, $E_{\rm ion}$, associated with the high-velocity dispersion component.
Assuming that $P_{\rm jet}$ represents the time-averaged jet power over $t_{\rm jet}$, we estimate the total energy injected by the jet during its propagation time as $E_{\rm jet} = P_{\rm jet} t_{\rm jet}$ \citep{Venturi2021_turmoil}.
The jet propagation time was estimated using Eq.~A1 from \cite{Mukherjee2018b}, adopting the same parameters as for the local Seyfert galaxy IC5063, to which their simulations were tailored, and assuming a jet radius equal to half of the largest linear size observable in the 1.4 GHz image of Fig.~\ref{fig: radio images}, corresponding to $r_\mathrm{jet}$ $\sim$ 0.8~kpc. We thus derive $t_{\rm jet} = 0.6$~Myr and $E_{\rm jet} = 2.7 \times 10^{55}$~erg. The ratio between the kinetic energy of the ionized gas and the total jet energy, $E_{\rm ion}/E_{\rm jet}$, is $\sim 10^{-3}$, in line with the samples of \cite{Venturi2021_turmoil} and \cite{Ulivi2024}. This result suggests that, even with a relatively low energy transfer efficiency, the jet in \ngc\ is capable of injecting the required energy into the ISM. In particular, it supports a scenario in which jet-driven feedback can operate effectively even in systems where the jet is oriented nearly perpendicular to the galaxy disk.

Interestingly, the observed correlations seem to also hold at higher jet-disk inclination angles, as in the case of \ngc, contrary to expectations from jet-ISM simulations, which predict a significantly weaker interaction and resulting kinematic perturbations before the jet breaks out of the disk \citep{Mukherjee2016, Mukherjee2018a, Mukherjee2018b, Meenakshi2022}. Although additional data are required to confirm this interpretation, it has important implications for studies lacking spatial resolution, where the jet inclination cannot be directly constrained, such as in high-redshift sources.
More generally, these correlations indicate that jets can efficiently inject energy into the ionized gas, driving perturbations across a wide range of radio powers. This mechanism may represent an important form of feedback, as the induced perturbations can inhibit gas cooling and suppress subsequent SF \citep{Mandal2021}.

\subsubsection{Nature of the $\sigma$-enhancement} \label{sec: nature sigma}
The origin of the $\sigma$-enhancement observed perpendicular to the radio jet in \ngc\ is suggestive of jet-ISM interaction, and is supported by several independent pieces of evidence, as the sufficient jet energy budget ($E_{\rm ion}/E_{\rm jet} \sim 10^{-3}$) and the agreement with the $P_{\rm jet}$-$M_{\rm ion}$ and $P_{\rm jet}$-$E_{\rm ion}$ correlations observed in other AGN host galaxies \citep[e.g., ][]{Mukherjee2018a, Jarvis2019, Venturi2021_turmoil, Meenakshi2022,Ulivi2024}. 
Enhancement of the velocity dispersion perpendicular to jets and ionization cones has been observed in several AGN hosting low-power jets at low inclination angles relative to the galaxy disk \citep[e.g.,][]{Venturi2021_turmoil,Ulivi2024}, and is predicted by simulations of jet–ISM interaction \citep{Mukherjee2016, Mukherjee2018a}. \ngc\ could represent a similar case, with the notable difference that here the jet is oriented perpendicular to the galactic disk plane rather than propagating within it. 
Supporting an ongoing interaction between the jet and the galaxy ISM more broadly, recent JWST/MIRI observations of \ngc\ reveal excess mid-IR dust heating on $\sim$100--200~pc scales in the NLR that requires shock-driven heating beyond AGN photoionization alone \citep{Haidar2026, Rosario2026}.

However, the jet-ISM interaction as interpretation for the perpendicular $\sigma$-enhancement is complicated by the fact that the stellar velocity dispersion shows a similar enhancement in the same region (see Fig.~\ref{fig: moment maps} and Sect.~\ref{sec: stellar kinematics}). 
We note that the spatial distribution of the gas $\sigma$-enhancement is co-spatial with the dust attenuation structure: where the narrow component is suppressed by dust, the broad component emerges and dominates the line profile, making the enhancement visible. This suggests that the gas $\sigma$-enhancement may be widespread in the entire galaxy disk, but only detectable where the dust geometry allows it. Remarkably, the stellar velocity dispersion follows the same spatial pattern, indicating that both components are likely tracing the same underlying dynamical perturbation -- regardless of the exact driving mechanism.
Since stars are collisionless and therefore not directly affected by jet-driven pressure waves on short timescales, the stellar broadening cannot be trivially explained by the same jet-ISM interaction mechanism \citep{Binney2008, Mukherjee2018a, Meenakshi2022}. This may suggest a pre-existing dynamical structure, as a bar or a kinematically hot component associated with the Virgo cluster \citep{Athanassoula1992, KormendyKennicutt2004, Cayatte1990}.
We finally note that the CO(2--1) kinematics traced by ALMA do not reveal any clear molecular counterpart to the perpendicular $\sigma$-enhancement, nor a molecular outflow component (Appendix~\ref{appendix: CO moment maps, PV diagram}). This suggests that, at the current angular resolution and sensitivity, cold molecular gas does not provide additional constraints on the physical origin of the ionized gas $\sigma$-enhancement.
Nevertheless, the CO(2--1) position-velocity diagram compared with the stellar velocity profile clearly highlights the presence and dynamical influence of the bar (see Fig.~\ref{fig: PV diagram}).
We therefore conclude that, while jet-ISM interaction remains a plausible driver of the gas $\sigma$-enhancement, the concomitant stellar signature prevents a definitive attribution. The nature of this kinematic feature remains ambiguous, and disentangling the contributions of jet feedback, gravitational perturbations, and  non-circular motions will require spatially resolved observations at sub-kpc scales.

\section{Summary and conclusions} \label{sec: summary}
In this work, we presented a detailed analysis of the ionized gas properties and kinematics in the central $\sim 5\times5$~kpc$^2$ of \ngc\ using MUSE WFM-AO observations, complemented by radio VLA data and archival ALMA CO(2--1) observations. This system provides an ideal laboratory to investigate the interplay between AGN-driven outflows, radio activity, and the host galaxy ISM, particularly in a configuration where the radio structure is oriented at high inclination with respect to the galaxy disk.
Our main results can be summarized as follows:

\begin{itemize}
    \item The MUSE data reveal a complex ionized gas structure in the central region of \ngc, where the gas kinematics show evidence of ordered disk motions and perturbed outflowing material. The \oiii\ emission clearly traces the ionization cones and the northeastern plume, while the \halpha\ mainly traces the disk and the star-forming regions (see Sect.~\ref{sec: moment maps}, Figs.~\ref{fig: rgb image} and \ref{fig: moment maps}).

    \item The stellar kinematics show a regular large-scale rotation pattern, with signatures of a non-axisymmetric structure, likely associated with a stellar bar and a nuclear disk. The stellar velocity dispersion map displays an asymmetric central enhancement that resembles the disturbed pattern observed in the ionized gas, suggesting that the inner kinematics is affected by dynamical perturbations beyond simple disk rotation (see Sect.~\ref{app: stellar fit}, Fig.~\ref{fig: moment maps}).

    \item The ionized gas moment maps reveal several distinct kinematic features. In addition to the rotating disk, we identify blueshifted gas in the northeastern plume and complex velocity substructures within both ionization cones. The \oiii\ velocity field, in particular, highlights clumpy approaching and receding regions that trace the ionized outflow (see Sect.~\ref{sec: moment maps}, Fig.~\ref{fig: moment maps}).

    \item The velocity dispersion maps show a clear $\sigma$-enhancement along the ionization cones, especially in the northern cone, with values up to $\sim$300~\kms, likely tracing the region where the outflow is still propagating within the galactic disk. In the southern cone, we identify a donut-shaped high-$\sigma$ structure, possibly associated with the approaching side of the outflow (see Sect.~\ref{sec: moment maps}, Fig.~\ref{fig: moment maps}).
    We detected an extended ionized-gas $\sigma$-enhancement within the galactic disk, roughly perpendicular to the radio/outflow axis. This feature is visible in all the main optical emission lines and is particularly strong in \oiii, where it extends up to $\sim$2 kpc. Its morphology and energetics are suggestive of being the result of jet-ISM interaction, though the cospatial enhancement of the stellar velocity dispersion prevents a definitive attribution to this mechanism alone (see Sect.~\ref{sec: nature sigma}, Figs.~\ref{fig: moment maps} and \ref{fig: w70}).


    \item The spatially resolved BPT and VO87 diagrams reveal a clear separation between star-forming, intermediate/LINER-like, and AGN-ionized regions. SF is mainly distributed along the spiral arms, AGN ionization dominates over the biconical outflow, and intermediate-excitation regions are found at the edges of the outflow and within the disk. In particular, the $\sigma$-enhanced regions perpendicular to the outflow axis show LINER/Composite-like line ratios, supporting a contribution from shock excitation associated with jet-ISM interaction (see Sect.~\ref{sec: attenuation and BPT}, Figs.~\ref{fig: bpt diagrams} and \ref{fig: bpt sigma}).
    Archival ALMA CO(2--1) data show only a marginal spatial correspondence between the molecular gas and the perpendicular $\sigma$-enhancement, with no clear evidence of a molecular outflow. Instead, the CO kinematics highlights the dynamical influence of the stellar bar on the gas motions in the central $\sim$2~kpc (see Appendix~\ref{appendix: CO moment maps, PV diagram}).

    \item Using the \moka\ 3D kinematic modeling, we reconstructed the intrinsic geometry and kinematics of the outflow, accounting simultaneously for the contribution of the rotating disk. From the \moka\ model and the attenuation-corrected emission-line luminosities, we derived the main properties of the ionized outflow. We estimate a mean outflow velocity of $\sim 400~\kms$, a total ionized outflow mass of $\sim 1.1\times10^6$~\msun, a mass outflow rate of $\sim 0.23$~\msun\,yr$^{-1}$, and a kinetic power of $\sim 2.8\times10^{40}$~erg\,s$^{-1}$. The northern and southern cones contribute comparably to the total energetics, despite their different densities and morphologies (see Sects.~\ref{sec: moka fit results} and \ref{app: moka modelling setup}, Fig.~\ref{fig: moka results}, Table~\ref{tab:moka_fit_outflow}).
    Comparing the outflow kinetic power with both the jet kinetic and the AGN radiative power, we find that either mechanism is energetically capable of driving the observed outflow, with a coupling efficiency of the order of 1$\%$ (see Sect.~\ref{sec: comparison outflow mechanisms}).

    \item We modeled the radial outflow velocity profiles with the energy-conserving outflow code \textsc{MAGNOFIT}. The models reproduce the observed acceleration and deceleration patterns, supporting an energy-conserving interpretation and indicating that the outflow 
    propagation is strongly shaped by the clumpy ISM density structure within the central few kpc (see Sect.~\ref{sec: vel profile model}, Appendix~\ref{appendix: setup profile model}, 
    Fig.~\ref{fig: vel profile model}).
    
    \item We compared the properties of the turbulent ionized gas with the kinetic power of the radio emission. The mass and kinetic energy of the gas associated with the $\sigma$-enhanced regions are consistent with the correlations found in other Seyfert galaxies, indicating that even relatively low-power radio jets can efficiently perturb the ISM and inject perturbations over kiloparsec scales (see Sect.~\ref{sec: comparison with literature}, Fig.~\ref{fig: plot energetics}).
    Finally, by comparing the kinetic energy of the disturbed ionized gas with the total energy released by the jet over its lifetime, we find that only a small fraction of the jet energy is required to explain the observed gas perturbations. This shows that the radio structure in \ngc\ is energetically capable of driving the observed feedback, even in a system where the radio axis is nearly perpendicular to the galaxy disk (see Sect.~\ref{sec: comparison with literature}).

\end{itemize}
Our results demonstrate that AGN-driven feedback in \ngc\ is not limited to the action of kpc-scale outflows along the ionization cones, but also includes widespread perturbations injected within the galactic disk. In particular, we find evidence that even radio structures oriented nearly perpendicular to the disk may significantly impact the host galaxy ISM, though the physical origin of the observed perturbations likely involves multiple concurring mechanisms. This implies that jet-ISM coupling, and the resulting turbulent feedback, should be considered a general and potentially important channel of AGN feedback, regardless of the relative orientation between the jet and the host galaxy disk.
Establishing the physical origin of this perpendicular $\sigma$-enhancement will require spatially resolved observations of larger samples spanning a broad range of jet powers and jet-disk inclinations, combined with multi-phase gas tracers and tailored simulations of jet-ISM interaction.

\begin{acknowledgements}
MC, EB, GC and AF acknowledge support from the INAF Fundamental 
Astrophysics programme 2024.
Based on observations No. 18B-163 and 19A-018 with the National Radio Astronomy Observatory and Green Bank Observatory which are facilities of the U.S. National Science Foundation operated under cooperative agreement by Associated Universities, Inc.
This paper makes use of the following ALMA data: ADS/JAO.ALMA\#2019.1.00763.L. ALMA is a partnership of ESO (representing its member states), NSF (USA) and NINS (Japan), together with NRC (Canada), NSTC and ASIAA (Taiwan), and KASI (Republic of Korea), in cooperation with the Republic of Chile. The Joint ALMA Observatory is operated by ESO, AUI/NRAO and NAOJ.
GV acknowledges support from European Union's HE ERC Starting Grant No. 101040227 - WINGS.
EA acknowledges financial support from INAF under the Mini Grant 2023 funding scheme (project ‘Low radio frequencies as a probe of AGN jet feedback at low and high redshift’).
MT and KZ are funded by the Research Council Lithuania grant no. S-MIP-24-100. ET acknowledges support from the ANID CATA-BASAL program FB210003, and FONDECYT Regular 1241005 and  1250821.
IL is supported by the European Research Council (ERC) under the European Union’s Horizon 2020 research and innovation programme (DistantDust, Grant agreement No. 101117541). MM acknowledges support from the European Research Council under ERC-AdG grant PICOGAL-101019746 and from the DFG Research Unit FOR-5195. 
\end{acknowledgements}

\bibliographystyle{aa}
\bibliography{biblio}

\clearpage

\begin{appendix}



\section{Observations and data reduction} \label{app: obs, data reduction}
\ngc \ was observed with MUSE \citep{Bacon2010} at the Very Large Telescope (VLT) in Wide Field Mode (WFM) with adaptive optics (AO) support during two nights, on 12 and 25 March 2022, under program 108.229J.001 (PI: G. Venturi).
The final reduced datacube combines six dithered and 90$^\circ$-rotated on-target exposures of 740~s each, acquired as part of two Observing Blocks (OBs). The total on-source exposure time is 4440~s and the FWHM effective spatial resolution is $\sim0.9''$.
For each OB, one sky exposure of 180~s was observed using the same instrumental configuration. 

We applied the standard MUSE data reduction pipeline (version 2.10.16), as extensively applied in previous works and described therein \citep[e.g.,][]{Venturi_2018, Mingozzi2019}.
One of the OBs was affected by spatially ubiquitous strong absorption features at observed wavelengths $\sim6475$--$6490$~\AA\ and $\sim6815$--$6835$~\AA.  Since the latter feature impacts the \siiALL doublet, the line fitting for these lines was performed using the datacube resulting from the unaffected OB only; for the rest of the spectral analysis, instead, the datacube obtained from both OBs was used, and the above spectral range was masked (see Sect.~\ref{app: data analysis} for details).

The final datacubes consist of $304 \times 296$ spaxels, corresponding to approximately $60.8 \times 59.2$ arcsec$^2$, given a MUSE spatial sampling of $0.2$ arcsec/spaxel. This field of view covers the central $\sim 5\times5$~kpc$^2$ of \ngc. The achieved spatial resolution is $\sim0.8$~arcsec ($\sim65$~pc). The MUSE spectral binning is $1.25$~\AA/channel, and its resolving power ($\lambda/\Delta\lambda$) in WFM is $1770$ at $4800$~\AA\ and $3590$ at $9300$~\AA.
Finally, we apply an astrometric correction, registering the position of the point source Gaia DR3 3907951233226666496 (RA~=~$12^{\rm h}25^{\rm m}47.57^{\rm s}$, Dec~=~$+12^\circ39'23.34''$), observable in our FoV.
The applied shifts are $\Delta$RA = +2.89\arcsec and $\Delta$Dec = $-$0.42\arcsec.

We analyzed archival \textit{Karl G. Jansky} Very Large Array (VLA) observations of \ngc\ in the C and K bands, obtained from projects 19A-018 and 18B-163 (PI: E. Chiaraluce), respectively.
The observations were carried out in March 2019 and in December 2018, with total integration times of 269~s and 449~s, covering the observed frequencies of 6 and 20 GHz, respectively.
We used the standard-calibrated visibilities provided by the NRAO VLA data archive\footnote{\url{https://science.nrao.edu/facilities/vla/archive/index}.} and processed the data using CASA version 6.6.1-17 \citep{casa}. For each dataset, we averaged all the spectral windows (spws), which are 32 in band C and 62 in band K. We generated the clean continuum maps using the \texttt{tclean} task and a Briggs weighting scheme with robust parameter $b =$ 0.5. We used the \texttt{hogbom} cleaning algorithm with a detection threshold of 3 times the rms sensitivity. To enable a direct comparison between the two maps, we applied a tapering (\texttt{uvtaper} keyword in the \texttt{tclean} task) to the K-band data in order to match the angular resolution of the C-band data.
The final continuum 6 and 20 GHz maps reach rms sensitivities of 0.025 mJy/beam and 0.014 mJy/beam, with synthesized beams of $1.30 \times 0.90$ arcsec$^2$ and $1.58\times1.22$ arcsec$^2$, corresponding to physical scales of $\sim110\times73$~pc$^2$ and $\sim130\times99$~pc$^2$.

We complemented the MUSE and VLA observations with Atacama Compact Array (ACA, 7-m array) Band~6 observations of \ngc\ (project 2019.1.00763.L; PI: Brown, Toby), targeting the CO(2--1) transition (rest frequency 230.538~GHz; observed at 228.613~GHz given the systemic velocity of \ngc). The observations were carried out on 22 December 2019 with 10 antennas, for a total on-source integration time of 14.878 minutes.
We started from the calibrated measurement sets produced by the standard ALMA calibration pipeline and processed the data with CASA (version 6.5.5.21; \citealp{casa}).
First, we subtracted the continuum emission from the spectral window containing the CO(2--1) line (spw~16) in the $uv$-plane using the \texttt{uvcontsub} task in CASA. The continuum was fitted with a first-order polynomial (\texttt{fitorder=1}) using the line-free channel ranges, combining spectral windows. 
We imaged the continuum-subtracted visibilities into a spectral-line cube using the \texttt{tclean} task with a mosaic gridder, Briggs weighting ($b=0.5$), and the \texttt{hogbom} deconvolution algorithm. We adopted a channel width of 1.953~MHz, corresponding to a velocity resolution of $\sim2.6$~km~s$^{-1}$, over 954 channels centered on the redshifted line frequency. Cleaning was performed automatically down to a threshold of 0.0945~Jy~beam$^{-1}$ using the \texttt{auto-multithresh} masking algorithm, corresponding to approximately $3\sigma$ for the spectral-line cube ($\sigma_{\rm cube}=0.0315$~Jy~beam$^{-1}$). 
The resulting datacube has a pixel scale of $0.87''$, an image size of $240\times108$ pixels , and a synthesized beam of $8.040''\times4.238''$ at PA~=~$77.141^\circ$.
We derived the integrated intensity (moment~0), intensity-weighted velocity (moment~1), and intensity-weighted velocity dispersion (moment~2) maps from the cleaned spectral-line cube using the \texttt{immoments} task over the channel range containing the line emission. To exclude low signal-to-noise pixels, we applied flux thresholds of 0.06~Jy~beam$^{-1}$ for the moment-0 map and 0.08~Jy~beam$^{-1}$ for the moment-1 and moment-2 maps.The moment-0 map of the CO(2--1) line reaches an rms sensitivity of 0.7106~Jy~beam$^{-1}$~km~s$^{-1}$.
A continuum image was produced separately from the same calibrated, non-continuum-subtracted visibilities, excluding the channel ranges containing line emission in the four spectral windows used (spws~16, 18, 20, and 22). We adopted the same imaging strategy as for the line cube, using \texttt{tclean} with a mosaic gridder, Briggs weighting ($b=0.5$), \texttt{hogbom} deconvolution, and \texttt{auto-multithresh} masking. The data were cleaned down to a threshold of 1.3~mJy~beam$^{-1}$. The resulting continuum map has an rms sensitivity of 0.4186~mJy~beam$^{-1}$ and a synthesized beam of $8.040''\times4.238''$ at PA~=~$77.141^\circ$.

\section{Data analysis} \label{app: data analysis}
For the data analysis, we followed the same methodology described in detail by \cite{Marasco2020, Tozzi2021, Ceci2025_6240}. Our procedure involved two consecutive spectral fits. First, we modeled the full datacube to spectrally isolate and remove the stellar continuum emission (see Sect. \ref{app: stellar fit}). Then, using the continuum-subtracted cube, we performed a second fit to model the gas emission lines (see Sect. \ref{app: lines fitting}), adopting a more flexible approach in the selection of line profiles and model constraints. 

\subsection{Modeling of the stellar continuum}
\label{app: stellar fit}
We began our analysis by modelling the stellar continuum. To achieve a sufficient signal-to-noise ratio (S/N) in each spatial element, we first applied a Voronoi adaptive binning scheme \citep{cappellari_adaptive_2003}, targeting an average S/N of at least 100 in the rest-frame wavelength range [8400~$\AA$, 8800~$\AA$]. This threshold ensures to accurate model the stellar continuum and its absorption features.
We then used the penalized PiXel-Fitting code (pPXF; \citealt{cappellari_parametric_2004}) to perform the spectral fitting. The method models the observed spectrum as a linear combination of stellar templates convolved with a Gaussian line-of-sight velocity distribution, simultaneously accounting for gas emission components. 
For the stellar population modelling, we adopted the E-MILES single stellar population templates \citep{Vazdekis2016}.
To take into account for possible spectral distorsions (e.g caused by dust extinction, imperfect flux calibration), we consider in the model also an additive third-degree polynomial. We built a model for the emission lines with a variable number of Gaussian components (from one to three), to account for different kinematic features along the LOS, and we added it to the previous continuum model, thus obtaining our total model. The kinematic parameters ($v$ and $\sigma$) of each Gaussian component are the same for all  emission lines. We fit each spectrum three times (with one, two, or three Gaussian components), and then applied a Kolmogorov-Smirnov (KS) test to the residuals to determine the optimal number of Gaussian components in each spaxel. Starting from a single component, we progressively increased the number of components until the residuals satisfied the condition p$\geq$0.5, which we adopted as the threshold for an acceptable fit \citep[see details in][]{Marasco2020,Ceci2025_6240}.
During this procedure, we masked the artifacts at observed wavelengths $\sim6475$--$6490$~\AA\ and $\sim6815$--$6835$~\AA\ (see Sect.~\ref{app: obs, data reduction}).

\subsection{Multi-Gaussian emission line fit}
\label{app: lines fitting}
After subtracting the stellar component, the resulting datacubes contain only the emission from the ionized gas. 
For the emission-line analysis, each spatial channel of the datacube is smoothed with a Gaussian kernel of width $\sigma_{\mathrm{smooth}} = 1$ pixel $(0.2\arcsec)$ to improve the S/N. 

We performed the emission-line fitting on a spaxel-by-spaxel basis, again adopting a multi-Gaussian component fitting (from one to three components) strategy to better reproduce the asymmetric line profiles. The kinematic parameters of each component are the velocity ($v$) and velocity dispersion ($\sigma$), where the latter is corrected for instrumental broadening, thus representing the intrinsic velocity dispersion.
To mitigate degeneracies in the most complex profiles, we tied the kinematic parameters of the first Gaussian component ($v$ and $\sigma$) to the stellar kinematics derived in the previous section. This constraint was applied only to spaxels belonging to the stellar disk, identified as the unmasked Voronoi bins in Fig.~\ref{fig: moment maps}.
To account for possible differences between gas and stellar kinematics, we allowed this component to vary within boundaries of $\Delta v$ = 50~\kms and $\Delta\sigma$ = 60~\kms, which we verified to be sufficient to encompass the gas component associated with the disk.
We fit the following emission lines: \hbeta, [\ion{O}{III}]$\lambda4959\AA$, \oiii, [\ion{N}{II}]$\lambda6548\AA$, \halpha, and [\ion{N}{II}]$\lambda6583\AA$. Due to its lower S/N, we only fit the amplitude of the \hbeta line and tie its kinematic parameters to those determined by the simultaneous fit of the other emission lines. We fixed the flux ratios of the [\ion{O}{III}]$\lambda\lambda4959,5007\AA$ and [\ion{N}{II}]$\lambda\lambda6548,6584\AA$ doublets to their theoretical value of 3.  
We performed the KS test again, changing the p-value this time. We increased it to 0.9 to reproduce the line profiles with higher accuracy.
Finally, in order to fit the \siiALL doublet, we used only the OB unaffected by the absorption artifacts (see Sect.~\ref{app: obs, data reduction}). Due to the lower S/N, these lines were fitted on a spaxel-by-spaxel basis, fixing the kinematic parameters to those derived from the other emission lines.
Finally, we fit the \siiALL doublet only in the OB unaffected by absorption artifacts (see Sect.~\ref{app: obs, data reduction}). Given their low S/N, we apply the same procedure used for \hbeta, that is we only fit their amplitudes on a spaxel-by-spaxel basis, setting their kinematic parameters to those derived for the other emission lines.

\section{W$_\text{70}$ map} \label{appendix: w70 map}
\begin{figure}
    \centering
    \includegraphics[width=.99\linewidth]{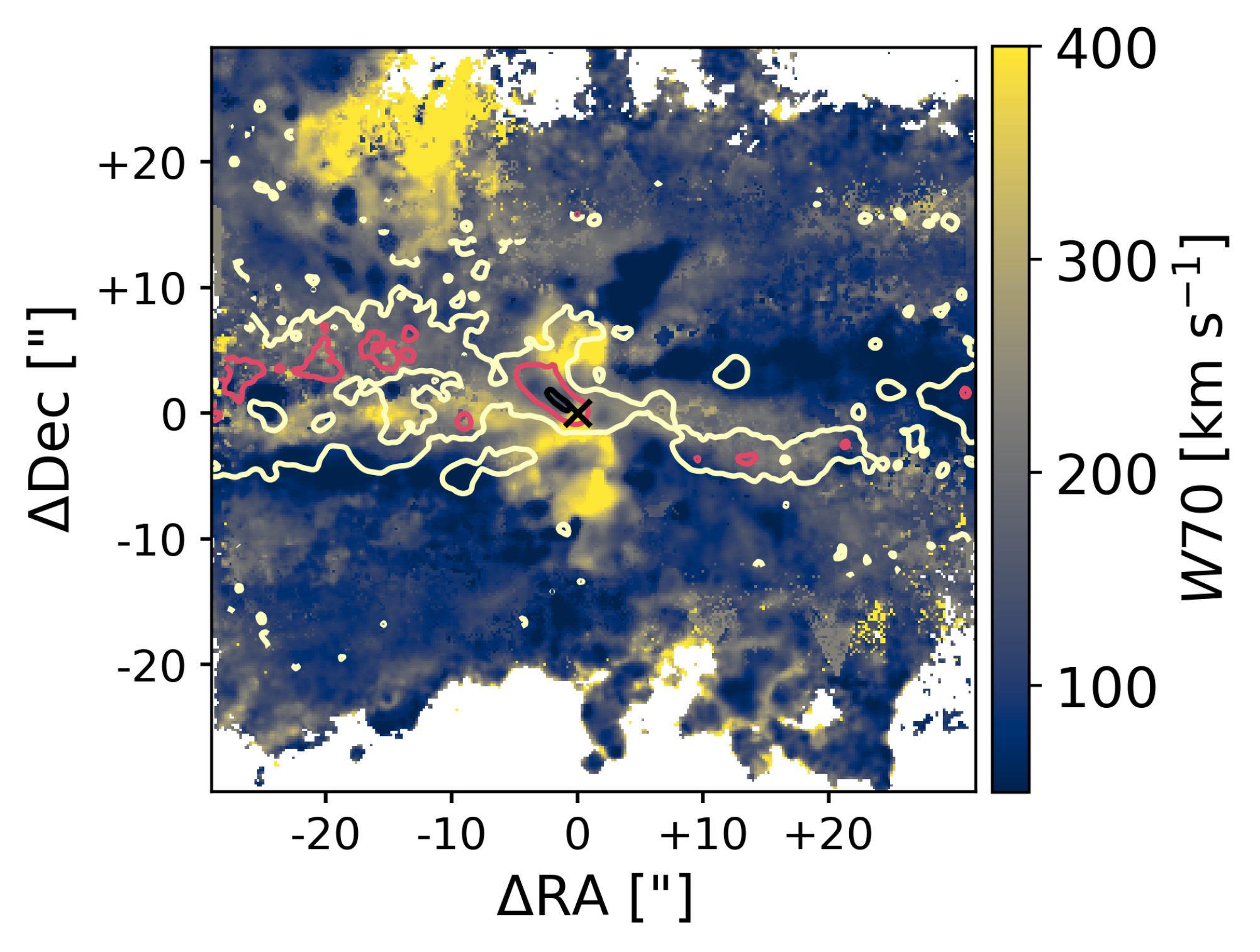}
    \caption{\oiii W$_{70}$ map. Contour levels correspond to $A_V = 2$, 3, and 4.5, shown in yellow, red, and black, respectively. The cross marks the position of the nucleus from the VLA 6\,GHz image.}
    \label{fig: w70}
\end{figure}

In this appendix, we present the \oiii W$_{70}$ map. In Fig.~\ref{fig: w70} we overlay $A_V$ contour levels on the W$_{70}$ map. Particularly within the galaxy disk, a correlation between high W$_{70}$ values and increased attenuation is evident. As discussed in the Sect.~\ref{sec: moment maps}, this argues against the interpretation that the observed $\sigma$-enhancement is driven by the superposition of multiple velocity components along the LOS.

\section{BPT and VO87 diagrams with $\sigma$ color-coding} \label{appendix: bpt sigma}
\begin{figure*}
    \centering
    \includegraphics[width=.8\linewidth]{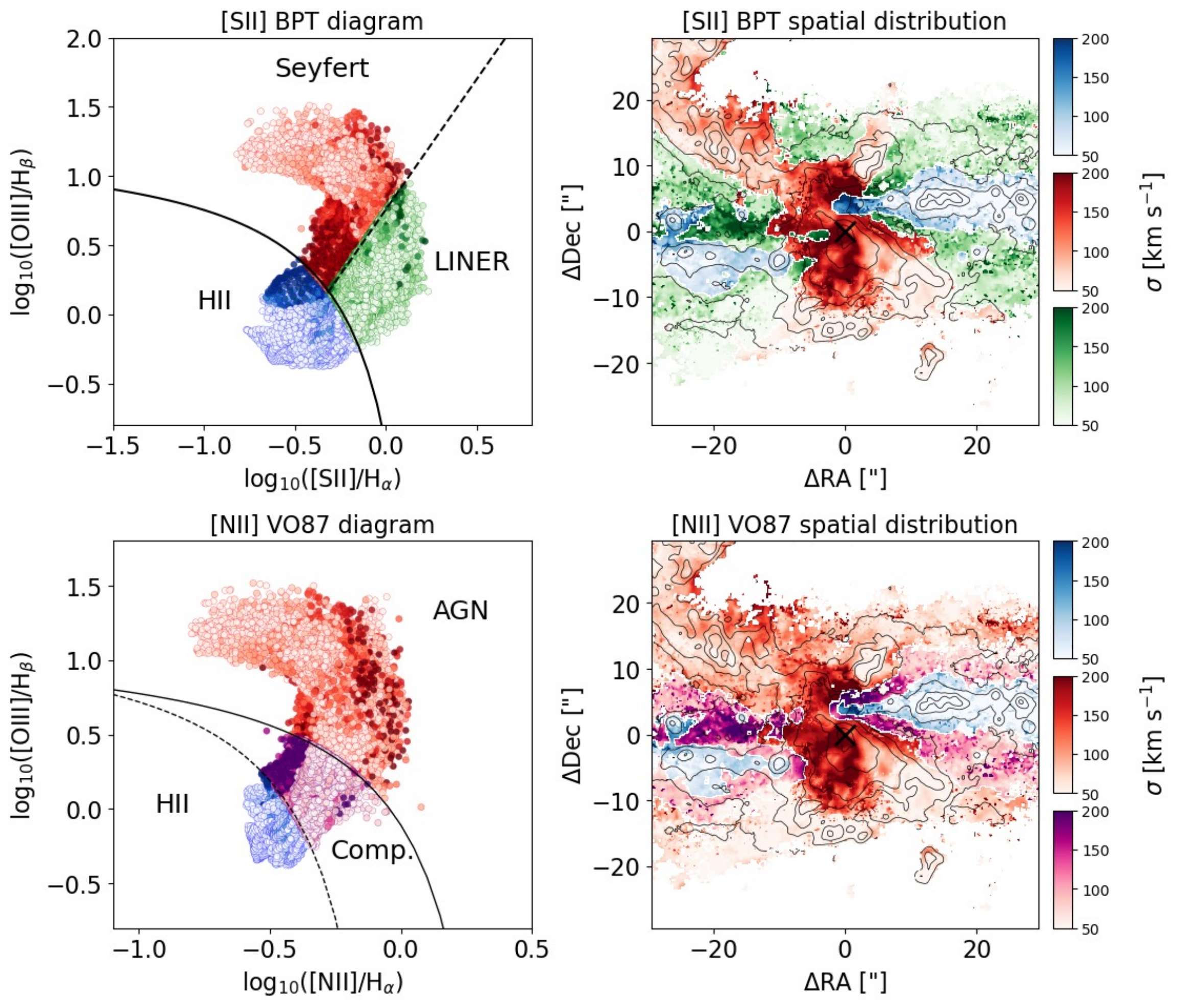}
    \caption{Same of Fig. \ref{fig: bpt diagrams}, but with the intensity of the colors scaled according with the velocity dispersion map of \oiii (see Fig. \ref{fig: moment maps}).} 
    \label{fig: bpt sigma}
\end{figure*}

In Fig.~\ref{fig: bpt sigma} we show the spatially resolved BPT and VO87 diagrams and the corresponding spatial distribution of the classified spaxels, color-coded by the velocity dispersion of the \oiii emission line. See Sect.~\ref{sec: attenuation and BPT} for a detailed discussion.

\section{\moka modeling setup} \label{app: moka modelling setup} 
As shown in the flux map of Fig.~\ref{fig: moment maps}, the morphology of the outflow is challenging to characterize due to its overlap with the galaxy disk in projection. Therefore, following the approach of \cite{Marconcini2025_nat}, we modeled the total \oiii spectral line profile by combining two kinematic components (outflow + disk) to best reproduce the observed features. 
We used the \oiii emission because it is the transition that best traces the outflow, as it is not blended with other high-S/N optical lines (e.g., \halpha\ and \nii) and is the warm-ionized transition with the highest S/N in the outflow component.
Specifically, we modeled the outflow and disk emission with a biconical outflow and a thin-disk model, respectively. In particular, we assumed a pure radial velocity field within the conical outflow and pure circular motions within the disk. Under the thin-disk assumption, we imposed a disk thickness equal to the PSF size of our observations ($\sim$ 1\arcsec, i.e. $\sim$ 80~pc). 
In our procedure, we fit the gas properties by dividing the conical outflow (disk) into concentric shells (annuli) of a fixed width of 1\arcsec, comparable to the minimum resolvable spatial scale of our observations.
The free model parameters in each shell are two: the outflow (disk) intrinsic radial (rotational) velocity and inclination with respect to the LOS. 

The fit procedure can be briefly summarized as follows. First, we masked the outflow region in the observed datacube from a visual inspection of the \oiii flux map in Fig.~\ref{fig: moment maps}, resulting in a 2D mask applied to the full datacube. We used a conical geometry with PA~=~20$^\circ$ and an opening angle of 110$^\circ$ (see Table~\ref{tab:moka_fit_outflow}). 
Then, we fitted the rotating gas disk emission in concentric annuli, leaving the rotational velocity and inclination of each annulus as free parameters. We stress that we imposed no constraints between the free parameters of adjacent shells, which are completely independent. This allowed us to reproduce the observed ionized emission in the galaxy disk. 
For simplicity, in our model we neglect both non-circular motions (e.g., bar-driven streaming; see \citealp{Veilleux1999a, Greene2014}) and radial variation of the disk PA, which we fix to PA~=~90$^\circ$, adopting the large-scale kinematic estimate from \cite{Veilleux1999a}. 
Next, we built a separate model cube for the outflow, assuming that the clouds follow a radial motion (see Eq. 3 in \citealt{Marconcini2023}), and combined it with the previously inferred best-fit disk model cube. As discussed in \cite{Marconcini2023}, to fit the outflow emission we fixed the axis position angle ($\gamma$), the conical aperture angle, and the center of the outflow model (assumed to coincide with the nucleus coordinates), since these parameters can be directly inferred from observations. Then, we fitted independently the outflow emission in both cones in each shell, simultaneously accounting for the underlying rotating disk. This was done by comparing the observed and modeled \oiii spectra shell by shell. 
The best-fit values for the disk and outflow parameters are reported in Table~\ref{tab:moka_fit_outflow}.

\section{CO(2--1) moment maps and position-velocity diagram} 
\label{appendix: CO moment maps, PV diagram}
\begin{figure*}
    \centering
    \includegraphics[width=1\linewidth]{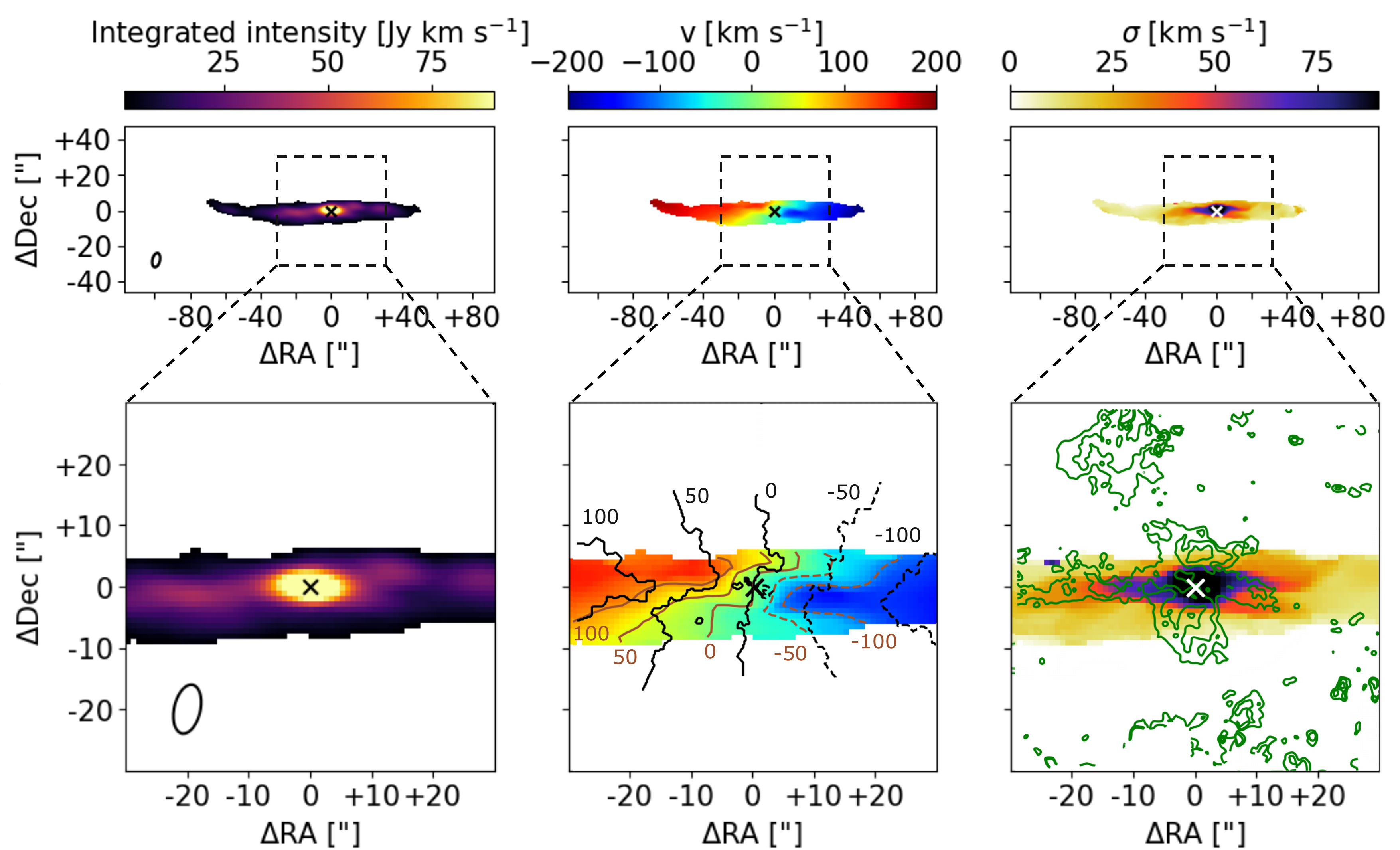}
    \caption{ALMA CO(2--1) moment maps of \ngc\ (top row) and corresponding MUSE FoV zoom-ins (bottom row). \textit{Top:} integrated intensity (zeroth-moment, left), intensity-weighted velocity (first-moment, center), and velocity dispersion (second-moment, right) of the CO(2--1) emission over the full ALMA field of view. The dashed black square marks the region covered by the MUSE observations. \textit{Bottom:} same CO(2--1) moments within the MUSE FoV. In the velocity panel (center), brown (black) contours trace the CO gas velocity (stellar LOS velocity field; see Fig.~\ref{fig: moment maps}) as indicated. In the dispersion panel (right), green contours show the \oiii\ second-moment map at (120, 150, 200, 250)~\kms\ (see Fig.~\ref{fig: moment maps}). The cross marks the position of the nucleus from the VLA 6\,GHz image, and the ellipse indicates the ALMA beam.}
    \label{fig: CO moment maps}
\end{figure*}
Here we present the CO(2--1) moment maps, in both the full ALMA FoV and the zoomed-in MUSE FoV  (Fig.~\ref{fig: CO moment maps}). 
At these wavelengths, dust attenuation is negligible. Therefore, the CO emission offers a more direct and reliable tracer of the gas distribution within the disk than optical tracers.
The flux map shows molecular gas emission in the disk, with an extension of $\sim120\arcsec$ ($\sim$10~kpc). In particular, the emission traces the two spiral arms also observed in the ionized component (see Fig.~\ref{fig: moment maps}). 
Moreover, the disk appears asymmetric, with an extension of $\sim50\arcsec$ to the W and $\sim70\arcsec$ to the E. This may be a signature of the ongoing or past ram pressure stripping event affecting \ngc\ as it moves through the Virgo cluster ICM \citep[e.g.,][]{Veilleux1999a, Verdugo2015, Vollmer2018}, consistent with the truncated HI and H$\alpha$ disk and the extraplanar gas plume reported in these studies.
The nuclear disk is unresolved at the ALMA angular resolution ($\sim8\arcsec\times4\arcsec$, see Sect.~\ref{app: obs, data reduction}) and we do not detect either any strong emission perpendicular to the galaxy disk or evidence of fast outflowing molecular gas.
In the first-moment map, we observe a rotating pattern distorted by the bar influence, with a maximum amplitude of $\sim180~\kms$, similar to the ionized counterpart (see Sect.~\ref{sec: moment maps}).  This is consistent with \citet{DominguezFernandez2020}, who analyses NOEMA CO(2--1) data at $0.4\arcsec-0.7\arcsec$ resolution and found that after subtracting  a rotating-disk mode, the residuals are consistent with bar-driven streaming motions in the plane of the galaxy, analogous to the inflow interpretation proposed by \citet{Veilleux1999a}. Due to the lower angular resolution of the ALMA data, however, we cannot resolve kinematic substructures as in the ionized gas. 
In Fig.~\ref{fig: CO moment maps}, bottom-central panel, we overplot both the CO(2--1) and the stellar velocity contours, with the same levels. The zero-velocity curve of the molecular gas is more strongly twisted than that of the stars, being more inclined with respect to the disk major axis. This indicates that the gaseous component is more affected by the bar potential and associated streaming motions. This behavior is consistent with the stronger response of the gas to non-axisymmetric bar torques compared to the collisionless stellar component (e.g. \citealp{Kim2024}).
The second-moment map roughly follows the flux distribution, with the highest values of $\sim100~\kms$ in the nucleus that decrease to $\sim10~\kms$ in the outer regions of the disk. Interestingly, the $\sigma$ enhancement across the galactic disk shows only a partial spatial correspondence with the ionized gas toward the eastern side.

\begin{figure}
    \centering
    \includegraphics[width=1\linewidth]{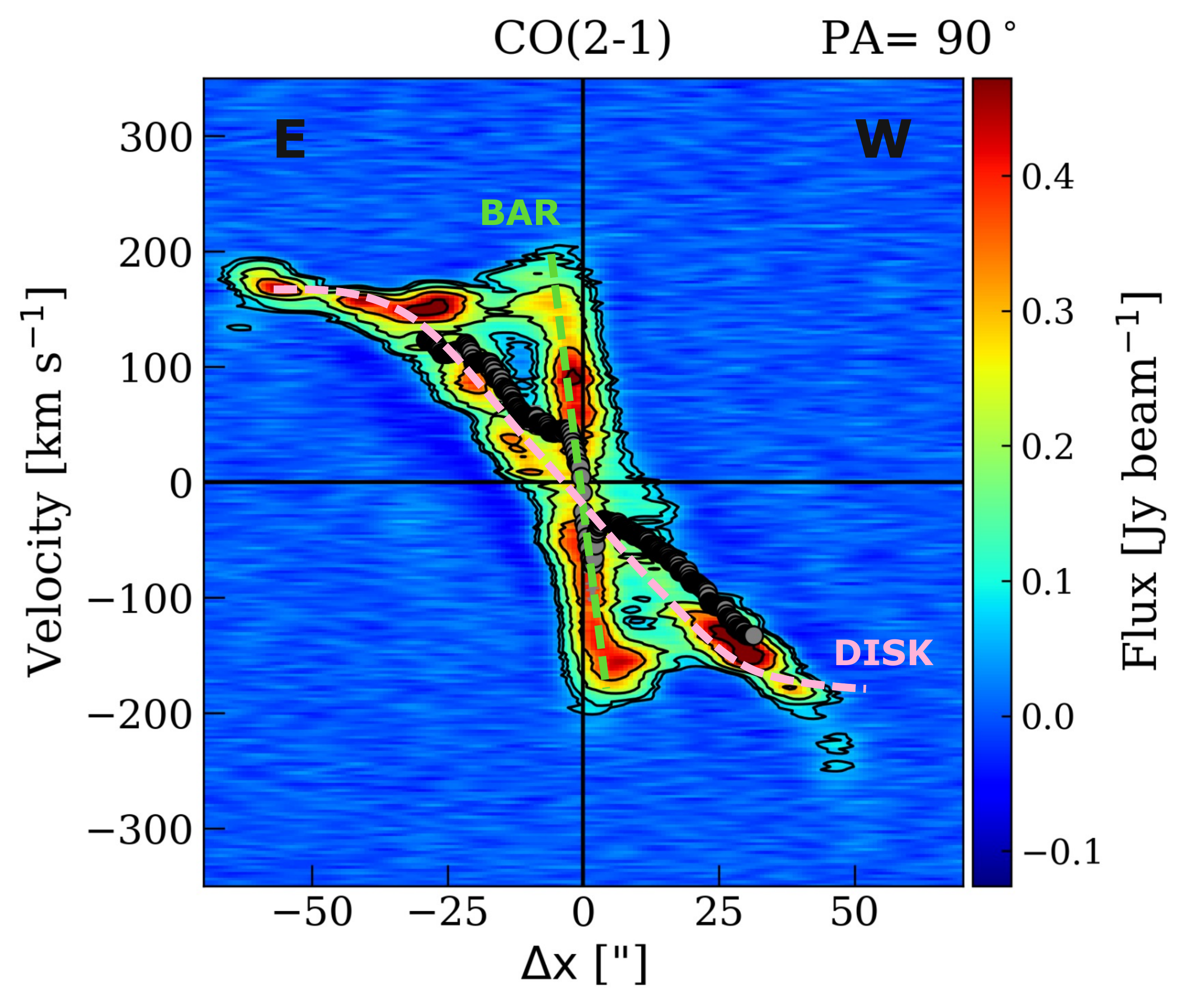}
    \caption{ CO(2--1) position-velocity diagram extracted along the major axis of the disk (PA~$\sim90^\circ$). The color scale and contours show the CO(2--1) flux distribution, while the overplotted grey circles trace the mean stellar velocity extracted within a $1\arcsec$-wide slit. The diagram reveals an X-shaped structure, with the smoothest branch (pink line) associated with the rotating disk and the steepest branch (green line) tracing the non-circular motions driven by the bar. Contours are at (2, 3, 6, 10, and 15$\sigma$), where $\sigma = 0.0315~\mathrm{Jy,beam^{-1}}$.}
    \label{fig: PV diagram}
\end{figure}

To better understand how the bar affects both the gas and stellar components, in Fig.~\ref{fig: PV diagram} we present the position-velocity (p-v) diagram extracted from the CO(2--1) data cube along the disk major axis (PA $\sim 90^\circ$) using a slit width of 6" in order to match the average beam size. The p-v diagram exhibits a clear X-shaped structure, which is characteristic of barred rotating disks \citep[e.g., ][]{Athanassoula&Bureau1999, Bureau&Athanassoula2005}. The maximum velocity amplitude reaches $\sim$200~\kms and remains nearly constant from the nucleus out to $\sim$50\arcsec ($\sim$4~kpc) from the center. 
\cite{DominguezFernandez2020} found an asymmetric emission in terms of flux, enhanced to the west, in the inner $20\arcsec$ p-v diagram, consistent with our findings. They attributed the non-circular motions in this region to the bar, likely driving the formation of a nuclear disk or ring.
We overplot the mean stellar velocity extracted within a $1\arcsec$-wide slit from the stellar velocity map (see Fig.~\ref{fig: moment maps}). The stellar velocities appear to trace the smoothest component of the p-v diagram. In the nuclear region ($\sim 5\arcsec$), the stars deviate from this trend, instead following the steepest features of the diagram, likely due to the bar influence. Probably we are observing the kinematics of the nuclear disk observable in Fig.~\ref{fig: attenuation map} (see \citealp{Greene2014, RodriguezArdila2017}). This behavior is also visible in the stellar velocity map (Fig.~\ref{fig: moment maps}), where higher velocity amplitudes are observed near the nucleus. Based on this, we suggest that the two structures forming the X-shape in the p-v diagram correspond to the disk, where stars are predominantly distributed, and the bar, which primarily influences the gas and, to a lesser extent, the stellar component, as indicated in Fig.~\ref{fig: PV diagram}.

\section{Outflow velocity profiles with MAGNOFIT}\label{appendix: setup profile model}
The propagation of galactic outflows could be explained by the simple AGN wind-driven outflow model \citep[for a review, see][]{King2015}. In this scenario, the AGN radiation field launches a quasi-relativistic wind from the accretion disk, carrying $\sim 5\%$ of the AGN bolometric luminosity in the form of kinetic power.
As the wind interacts with the surrounding ISM, it produces a shock that heats the gas to temperatures of $\sim 10^{10}$~K, driving the expansion of a large-scale outflow. In most theoretical models, the shocked wind bubble expands approximately adiabatically, such that the outflow is energy-driven, i.e., powered by the total energy input of the AGN wind \citep[e.g.][]{Faucher-Giguere2012}. This results in high outflow velocities, typically $v_{\rm out} \gg \sigma_{\rm bulge}$.
On larger scales, the outflow evolution depends on the ambient gas density profile. For an isothermal distribution ($\rho \propto R^{-2}$), the outflow rapidly reaches a constant velocity. Shallower density profiles lead to a deceleration, whereas steeper profiles produce accelerating outflows. 
Once the outflow propagates beyond the dense central region into the lower-density halo, the reduced ambient pressure allows the bubble to expand more freely, potentially leading to outflow acceleration at larger radii.

The numerical scheme \textsc{MAGNOFIT}, used in Sect.~\ref{sec: vel profile model}, is a 1D integrator that follows the evolution of the outflow radius and velocity within a spherically symmetric mass distribution composed of components with analytically defined mass profiles. 
The underlying equation of motion, described in detail in the Appendix of \cite{Zubovas2022}, generalizes the formulation of energy-conserving outflows presented by \cite{King2015}.
More recently, this code has been applied to interpret the acceleration of outflows in a sample of galaxies with spatially resolved ionized gas kinematics \citep{Zubovas2025}. In that context, the observed acceleration can be reproduced with a two-component model consisting of a gas-rich bulge embedded within a gas-poor halo, where the outflow accelerates after reaching the edge of the bulge. In \cite{Marconcini2025}, the same framework was applied to NGC~1068, showing that its more complex outflow kinematics can be reproduced with a multi-component bulge-halo system.

We fixed the SMBH mass to $8.5 \times 10^6\ \msun$, as derived from water maser emission by \cite{Kuo2011}. We then estimated a velocity dispersion of $\sigma \sim 111.6$~\kms\ using the $M-\sigma$ relation from \citet{McConnell2013}. We fixed the AGN luminosity to $L = 0.5 L_{\rm Edd} = 5.36 \times 10^{44}$~erg~s$^{-1}$, which exceeds the current luminosity estimate by about a factor of six; however, since AGN luminosity can change on timescales much shorter than the outflow dynamical time, the present-day luminosity is not necessarily representative of the long-term average. Additionally, if the luminosity is scaled in proportion with the gas fraction (see below), the outflow expansion remains the same \citep{Zubovas2025}.
We applied the fit independently to the two outflow velocity profiles, adopting a multi-component bulge-halo model. In the case of the southern outflow (lower panel in Fig.~\ref{fig: vel profile model}), a high-density, narrow component at $R \sim 0.8$~kpc is required to reproduce the observed deceleration. In contrast, for the northern outflow (middle panel in Fig.~\ref{fig: vel profile model}), a broader shell combined with a low-density gap reproduces both the acceleration up to $R \sim 0.5$~kpc and the subsequent sharp deceleration.
The combined fit, as expected, recovers similar structures, namely a low-density gap together with both a broader and a narrower obstructing shell (upper panel in Fig.~\ref{fig: vel profile model}).
Interestingly, all fits independently converge to consistent values of the total bulge mass, $M_{\mathrm{bulge}} \sim 6.51 \times 10^9\ \msun$. This value is comparable with previous literature estimates based on photometric decompositions, such as $\log(M_{\ast,\mathrm{sph}}/\msun) \simeq 9.77$ from \citet{Lasker2016} and  $\log(M_{\ast,\mathrm{sph}}/\msun) \simeq 10.07$ from \citet{Davis2019}. 
Given the uncertainties associated with decomposing the barred, highly inclined stellar structure of \ngc, and the fact that the \textsc{MAGNOFIT} parameter represents an effective mass component governing the outflow propagation, this level of agreement is reasonable.

\section{$P_{\text{jet}}$ from VLA data} \label{appendix: Pjet}
\begin{figure*}
    \centering
    \includegraphics[width=1\linewidth]{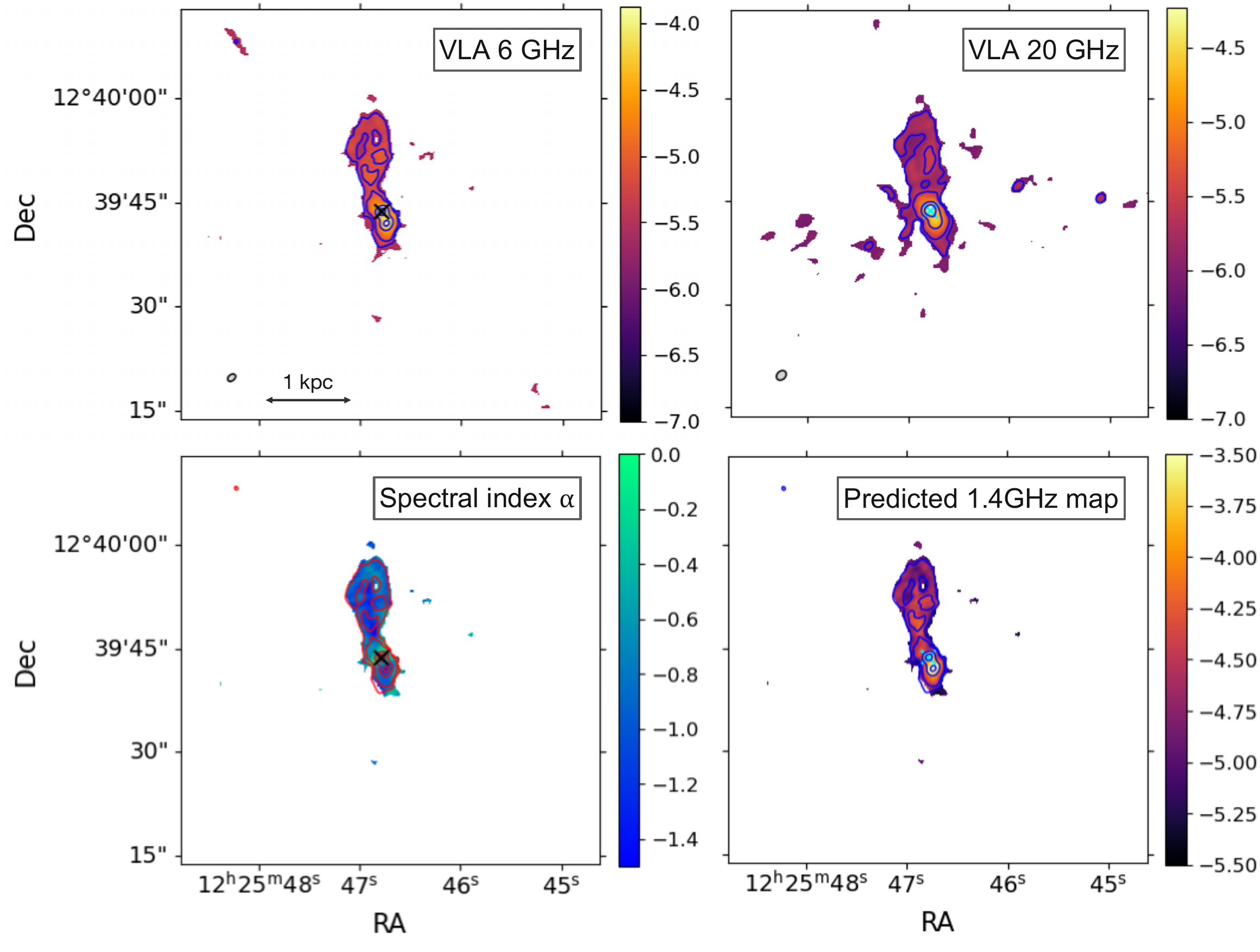}
    \caption{VLA radio continuum spectral flux density and spectral index maps in the same FoV as the MUSE data (see Fig.~\ref{fig: rgb image}). 
    \textit{Upper panels}: observed VLA images at 6~GHz (left) and 20~GHz (right). The VLA beams are shown as grey ovals in the bottom-left corner.
    \textit{Bottom-left}: spectral index map. 
    \textit{Bottom-right}: predicted 1.4~GHz continuum obtained by extrapolating the 6~GHz map using the spectral index map.
    Contours in the 6~GHz, 1.4~GHz and spectral index map show the $(5,\,10,\,50,\,120)~\times~$rms flux levels of the VLA 6~GHz image.
    Contours in the 20~GHz map show the $(5, 10,\,50,\,120)~\times~$rms flux levels.    
    The cross marks the position of the nucleus from the VLA 6~GHz map.
    The radio continuum maps are in units of Jy pixel$^{-1}$.
    Spaxels with emission below $3 \times \mathrm{rms}$ are masked.   
    }
    \label{fig: radio images}
\end{figure*}

In this appendix, we derive the radio luminosity at 1.4 GHz ($P_{\text{1.4\,GHz}}$) used to estimate the jet kinetic power in Sect.~\ref{sec: comparison with literature}. 
In the upper panels of Fig.~\ref{fig: radio images}, we present the VLA images at 6 and 20 GHz, both showing the same FoV as the MUSE data. The 20 GHz image was first beam matched to the 6~GHz image and then reprojected onto the same pixel grid, to ensure a consistent spatial sampling.
We then computed the spectral index $\alpha$ assuming $F_\nu \propto \nu^{\alpha}$, using
\begin{equation}
\alpha = \frac{\log(F_{\text{6\,GHz}}) - \log(F_{\text{20\,GHz}})}{\log(6) - \log(20)},
\end{equation}
as shown in the bottom-left panel of Fig.~\ref{fig: radio images}.
Finally, we extrapolated the 6 GHz map to 1.4 GHz by applying the derived spectral index on a pixel-by-pixel basis, obtaining the predicted 1.4 GHz map (bottom-right panel of Fig.~\ref{fig: radio images}).

\end{appendix}

\end{document}